\documentstyle[amssymb,prb,aps,tighten,preprint,eqsecnum]{revtex}

\begin{document}
\draft
\title{Coulomb effects in granular materials at not very low temperatures. }
\author{K.B. Efetov$^{(1,2)},$ A. Tschersich$^{\left( 1\right) }$}
\address{$^{(1)}$Theoretische Physik III,\\
Ruhr-Universit\"{a}t Bochum, 44780 Bochum, Germany\\
$^{(2)}$L.D. Landau Institute for Theoretical Physics, 117940 Moscow,\\
Russia \\
}
\date{\today}
\maketitle

\begin{abstract}
We consider effects of Coulomb interaction in a granular normal metal at not
very low temperatures suppressing weak localization effects. In this limit
calculations with the initial electron Hamiltonian are reduced to
integrations over a phase variable with an effective action, which can be
considered as a bosonization for the granular metal. Conditions of the
applicability of the effective action are considered in detail and
importance of winding numbers for the phase variables is emphasized.
Explicit calculations are carried out for the conductivity and the tunneling
density of states in the limits of large $g\gg 1$ and small $g\ll 1$
tunnelling conductances. It is demonstrated for any dimension of the array
of the grains that at small $g$ the conductivity and the tunnelling density
of states decay with temperature exponentially. At large $g$ the
conductivity also decays with decreasing the temparature and its temperature
dependence is logarithmic independent of dimensionality and presence of a
magnetic field. The tunnelling density of states for $g\gg 1$ is anomalous
in any dimension but the anomaly is stronger than logarithmic in low
dimensions and is similar to that for disordered systems. The formulae
derived are compared with existing experiments. The logarithmic behavior of
the conductivity at large $g$ obtained in our model can explain numerous
experiments on systems with a granular structure including some high $T_{c}$
materials.
\end{abstract}

\pacs{PACS numbers: 73.23.Hk, 73.22.Lp, 71.30.+h}


\section{Introduction}

Study of disordered systems attracts a lot of attention. Even if the
electron-electron interaction is neglected the problem is not simple.
Nevertheless, by now a rather complete understanding for non-interacting
systems exists. It is well known that systems with one-dimensional geometry
must be insulators for any week disorder. Two-dimensional systems are also
believed to be localized, whereas there should occur the Anderson
metal-insulator transition in three-dimensional disordered systems. This
scenario has been formulated on the basis of scaling arguments long ago \cite
{abrahams} and has been checked by numerous methods including diagrammatic
expansions \cite{gorkov} and non-linear $\sigma $-model calculations \cite
{wegner} (for a review, see e.g. Ref. \cite{efetov}).

At weak disorder, one obtains so called weak localization corrections which
are divergent in $1D$ and $2D$ (in $2D$ they are logarithmic). The infrared
divergency is cut off either by the sample size or external frequency, or by
the inverse inelastic time. Within the scaling theory the concrete mechanism
of this cutoff does not matter. Due to the existence of the cutoffs flow
diagrams stop at some point when changing parameters like the sample size or
temperature. So, within this picture by increasing the temperature one
cannot expect anything except cutting the weak localization corrections.

If an electron-electron interaction is added additional corrections
discovered by Altshuler and Aronov\cite{altar} become important. They are
also logarithmically divergent in $2D$ but remain convergent in $3D.$ For
the systems with the electron-electron interaction one can also derive a
proper $\sigma $-model \cite{fin} and demonstrate the renormalizability.
Within such an approach the role played by the temperature is, as for the
non-interacting models, to cut the diverging infrared corrections. After
they are cut off the system is assumed to become a conventional normal
metal. Although granular systems with the electron-electron interaction were
not considered explicitly, it is generally expected that there should not be
a qualitative difference between them and ``just disordered'' materials.
This is definitely true for systems without the electron-electron
interaction.

Therefore, a recent experimental observation \cite{Gerber97} came as a
surprise. In the experiment \cite{Gerber97} measurements of conductivity of
granular $Al-Ge$ thick films were performed and several interesting effects
have been discovered. These films consisted of $Al$ grains embedded in an
amorphous $Ge$ matrix. The size of an $Al$ grain was about $120$\AA\ and the
grains were at low temperatures in a superconducting state.

Destroying the superconducting pairing by a magnetic field up to $17T$ the
authors could study, in particular, properties of the normal state. Some
features of the normal state related to a negative magnetoresistance due to
superconducting fluctuations have been discussed recently \cite{bel} but
another unusual observation remained unexplained.

The authors of Ref. \cite{Gerber97} found in some samples a peculiar
temperature dependence of the conductivity. Samples that had a high room
temperature resistivity (this corresponds to a weak tunneling between the
grains) showed an exponential decay of the conductivity as a function of
temperature. This behavior is typical for insulators and has been
interpreted in Ref. \cite{Gerber97} in this way. Samples with larger
intergranular couplings did not show any exponential decay but the
resistivity did not saturate at low temperatures and the authors described
its temperature dependence by a power law 
\begin{equation}
R=AT^{-\alpha },  \label{a1}
\end{equation}
with $\alpha =0.117$. Apparently, with such a small value of $\alpha $ a
logarithmic temperature dependence 
\begin{equation}
R=A\left( 1-\alpha \ln T\right)  \label{a1a}
\end{equation}
(obtained by expansion of Eq. (\ref{a1}) in $\alpha )$ could describe the
experimental data as well.

Both the power law dependence, Eq. (\ref{a1}), and the logarithmic
dependence, Eq. (\ref{a1a}), were not expected from the theories of
disordered metals. Although the logarithmic dependence is natural of $2D$
films, it is relevant to emphasize that the array of the grains was three
dimensional and the behavior described by Eqs. (\ref{a1a}) was observed at
high magnetic fields. This definitely excludes an interpretation of the
logarithmic or power law behavior in terms of weak localization corrections
typical for $2D$ \cite{gorkov,altar,fin}.

In this paper we try to explain the transition from the exponential
temperature behavior at small $g$ to the logarithmic dependence of Eq. (\ref
{a1a}) assuming that the temperature is not very low, which actually
corresponds to the experiment \cite{Gerber97} where the lowest temperature
was around $0.3K$. We consider a model of granular metals at not very low
temperature and demonstrate that changing the dimensionless tunneling
conductance $g$ one can have either exponential temperature dependence of
the resistivity at small $g\lesssim 1$ or the logarithmic behavior, Eq. (\ref
{a1a}), at large $g\gtrsim 1$. It will be shown that the result is
applicable for any dimensionality of the array of grains, which contrasts
usual logarithmic corrections due to interference effects typical for $2D$
only.

Study of the regime of not very low temperatures of the granular metals has
started recently in Ref. \cite{belalt} where it was demonstrated that the
regime of the temperatures 
\begin{equation}
T\gg g\delta ,  \label{ad1}
\end{equation}
where $\delta $ is the mean level spacing, is clearly different from the
regime of lower temperatures. In this regime, well known weak localization
effects are suppressed but, nevertheless, the system may exhibit a
non-trivial behavior.

Eq. (\ref{ad1}) is written for $g\gg 1$. In this limit, one can study the
model by summation of diagrams containing impurity lines, as it has been
suggested in Ref. \cite{belalt}. Non-trivial effects arise from a
renormalization of the Coulomb interaction by impurities. Actually, a
non-trivial behavior originates from the diagrams of the Altshuler-Aronov
type \cite{altar} modified for the granular system. Therefore, for large $%
g\gg 1$ the theory presented in this paper is equivalent to the summation of
the Altshuler -Aronov diagrams. However we use an effective phase
functional, which gives us an opportunity to consider not only the first
order in the Coulomb interaction but to go into higher orders by writing
renormalization group equations.

If the tunneling conductance is not very large, $g\lesssim 1$, the
inequality (\ref{ad1}) should be replaced by the inequality 
\begin{equation}
T\gg \delta  \label{ad2}
\end{equation}
In the limit $g\lesssim 1$, it is not possible to get any results by
summation of conventional diagrams and one should use the phase functional.
A very important result of Ref. \cite{belalt} is that in the limit of not
very low temperatures, Eqs. (\ref{ad1},\ref{ad2}), the phase functional has
a form similar to that suggested by Ambegaokar, Eckern and Sch\"{o}n (AES)
long ago \cite{aes} for a description of quantum dissipation \cite{leggett}
in metals.

At lower temperatures $T\lesssim g\delta $ one must take into account low
energy diffusion modes, and the system should be described by a more
complicated non-linear $\sigma $-model containing both the phases $\phi $
and matrices $Q$. Of course, in the limit $T\rightarrow 0$ the AES
functional is no longer applicable and the notion of ``quantum dissipation''
looses its sense. In other words, the ``quantum dissipation'' corresponds to
a limit of not very low temperatures for the system under consideration.

It is relevant to mention that the model of a granular metal may roughly
describe disordered systems with a low electron concentration. One can
imagine that in such systems electrons spend a long time in some kind of
traps or puddles and tunnel between them with a small rate. Then, such a
situation resembles the granular systems.

A short account of the ideas of the present work has been presented in a
recent paper\cite{et}

The article is organized as follows:

In Sec. II we formulate the model and derive a free energy functional
containing phases. We discuss how one should integrate over the phases
accounting for winding numbers. In Sec. III we consider conductivity in the
limit of large conductances $g\gg 1$ and obtain the temperature behavior,
Eq. (\ref{a1a}). For calculations we use both the perturbation theory and
renormalization group techniques. In Sec. IV we consider the tunneling
density of states in the limit $g\gg 1$. The final result for a two
dimensional array of grains, Eq. (\ref{c8}), has the same form as the
corresponding formula for a disordered ``homogeneous'' metal. In Sec. V we
consider the limit of small $g\ll 1$ and show how one can carry out
summation over winding numbers. Physical quantities like the conductivity
and the tunneling density of states are shown to be exponentially small in
temperature. In Sec. VI we discuss the results and make a comparison with
the experiment.

\section{Choice of the model. Phase functional.}

In the present work, we consider a simplified model where metal grains form
a regular lattice. The grains are disordered due to impurities or irregular
boundaries. Each grain is separated from its nearest neighbors by tunneling
barriers. We assume that the main contribution to the macroscopic
resistivity of the granular system comes from the intergranular tunneling.

The Hamiltonian describing the model is chosen as 
\begin{equation}
\hat{H}=\hat{H}_{0}+\hat{H}_{t}+\hat{H}_{c},  \label{a2}
\end{equation}
where $\hat{H}_{0}$ is the one-electron Hamiltonian of isolated grains
including disorder within the grains 
\[
\hat{H}_{0}=\int \psi ^{+}\left( {\bf r}\right) \left( -\frac{{\bf \nabla }%
^{2}}{2m}+U\left( {\bf r}\right) \right) \psi \left( {\bf r}\right) d{\bf r},
\]
where $U\left( {\bf r}\right) $ is a random potential.

The tunneling of the electrons between the grains is given by 
\begin{equation}
\hat{H}_{t}=\sum_{{\bf i},{\bf j},\alpha ,\alpha ^{\prime }}t_{{\bf ij}}\hat{%
\psi}_{\alpha {\bf i}}^{+}\hat{\psi}_{\alpha ^{\prime }{\bf j}}.  \label{a2a}
\end{equation}
where the summation is performed over the states $\alpha $, $\alpha ^{\prime
}$ of each grain (spin is conserved) and over neighboring grains $i$ and $j$%
. The possibility to tunnel from the state $\alpha $ to an arbitrary state $%
\alpha ^{\prime }$ of other grains introduces an additional disorder
resulting in a finite tunnel conductance.

The term $\hat{H}_{c}$ in Eq. (\ref{a2}) describes the charging energy 
\begin{equation}
\hat{H}_{c}=\frac{e^{2}}{2}\sum_{{\bf ij}}\hat{N}_{{\bf i}}C_{{\bf ij}}^{-1}%
\hat{N}_{{\bf j}}.  \label{a3}
\end{equation}
In Eq. (\ref{a3}), 
\begin{equation}
\hat{N}_{{\bf i}}=\int \hat{\psi}^{+}\left( {\bf r}_{{\bf i}}\right) \hat{%
\psi}\left( {\bf r}_{{\bf i}}\right) d{\bf r}_{{\bf i}}-\bar{N}  \label{a3a}
\end{equation}
is the excess number of electrons in the ${\bf i}$-th grain, $\bar{N}$ is
the dimensionless local potential and $C_{{\bf ij}}$ is the capacitance
matrix. The integration over ${\bf r}$ in Eqs. (\ref{a2}, \ref{a3a}) is
performed over the grain with the coordinate ${\bf i}$ of the center and
includes summation over the spin.

Eq. (\ref{a3}) describes the long range part of the Coulomb interaction in
the limit of weak disorder inside the grains and has been used in many works
(for a review, see \cite{abg}). In principle, one could consider also a
superconducting or magnetic part of the interaction within the dot but we
assume that the grains are in the normal state. The long range part of the
Coulomb interaction, Eq. (\ref{a3}), describes the classical charging
energy. This interaction can lead to the Coulomb blockade and to insulating
macroscopic properties of the system of the grains.

Calculations with the Hamiltonian $\hat{H}$, Eqs. (\ref{a2}, \ref{a3}), can
be replaced in a standard way by computation of a functional integral over
anticommuting fields $\psi \left( {\bf r}_{{\bf i}}{\bf ,}\tau \right) $.

Although the model described by Eqs. (\ref{a2}, \ref{a3}) contains only the
long range part of the Coulomb interaction, it is still very complicated,
because at very low temperatures interference becomes very important and one
has to consider an interplay of localization and interaction effects. One
could do this either using diagrammatic expansions \cite{altar} or writing a
non-linear $\sigma $-model \cite{fin}. Both methods allow to consider the
limit of large tunneling conductances $g$ and the results are strongly
dependent on the dimensionality. However, the behavior, Eq. (\ref{a1}) or
Eq. (\ref{a1a}) was not predicted for $3D$ in any of these works.

The model, Eqs. (\ref{a2}, \ref{a3}), becomes simpler if the temperature $T$
is not very low such that low energy diffusion modes are damped. As it was
discussed in a recent publication \cite{belalt}, the granular metal can be
well described at temperatures $T\gg g\delta $, where $\delta $ is the mean
level spacing in a single grain, by the Ambegaokar, Eckern and Sch\"{o}n
(AES)\cite{aes} functional of the free energy. If $g\lesssim 1$, this
condition should be replaced by $T\gg \delta $. The limit of not very low
temperatures not only simplifies the consideration but is interesting on its
own because it leads to an unusual behavior of physical quantities and is
easily accessible experimentally. In particular, we will see that changing
the tunneling conductance $g$ one may have a crossover from the exponential
temperature dependence of the resistivity to the logarithmic behavior, Eq. (%
\ref{a1a}).

We calculate the conductivity $\sigma \left( \omega \right) $ using the Kubo
formula and making an analytical continuation from Matsubara frequencies $%
i\Omega _{n}=2\pi inT$ to real frequencies $\omega $ \cite{AGD}. Within this
formalism the conductivity $\sigma \left( \omega \right) $ can be written in
the form 
\begin{equation}
\sigma \left( \omega \right) =\frac{ia^{d-2}}{\omega }\left[ \sum_{{\bf j}%
}\int d{\bf r}_{{\bf j}}\int_{0}^{\beta }d\tau e^{i\Omega _{n}\tau }K_{{\bf %
aa}}\left( {\bf r}_{{\bf i}}-{\bf r}_{{\bf j}},\tau \right) \right] _{\Omega
_{n}\rightarrow -i\omega +\delta }  \label{b1}
\end{equation}
\bigskip where ${\bf a}$ is a vector connecting the centers of neighboring
grains ${\bf i}$ and ${\bf i}+{\bf a}$, $a=\left| {\bf a}\right| ,$ and $d$
is the dimensionality of the array.

The function $K_{{\bf aa}^{\prime }}$ in Eq. (\ref{b1}) can be written as 
\begin{equation}
K_{{\bf aa}^{\prime }}\left( {\bf r}_{{\bf i}},{\bf r}_{{\bf j}};\tau
\right) =\Pi _{{\bf aa}^{\prime }}\left( {\bf r}_{i},{\bf r}_{j};\tau
\right) -e^{2}\delta _{{\bf aa}^{\prime }}\delta _{{\bf ij}}\delta \left(
\tau \right) <\hat{H}_{t{\bf i}}^{{\bf a}}\left( \tau \right) +h.c.>
\label{b2}
\end{equation}
where 
\begin{equation}
\Pi _{{\bf aa}^{\prime }}\left( {\bf r}_{{\bf i}},{\bf r}_{{\bf j}};\tau
\right) =-<T_{\tau }\hat{J}_{{\bf a}}\left( {\bf r}_{{\bf i}},\tau \right) 
\hat{J}_{{\bf a}^{\prime }}\left( {\bf r}_{{\bf j}},0\right) >  \label{b3}
\end{equation}
and 
\begin{equation}
\hat{H}_{t{\bf i}}^{{\bf a}}\left( \tau \right) =\sum_{\alpha ,\alpha
^{\prime }}t_{{\bf i+a,i}}\hat{\psi}_{\alpha ,{\bf i+a}}^{+}\left( \tau
\right) \hat{\psi}_{\alpha ^{\prime },{\bf i}}\left( \tau \right)  \label{b4}
\end{equation}

The abbreviation $h.c.$ means a Hermitian conjugation. The tunneling current
operator $\hat{J}_{{\bf a}}\left( {\bf r}_{{\bf i}},\tau \right) $ entering
Eq. (\ref{b3}) takes the standard form 
\begin{equation}
\hat{J}_{{\bf a}}\left( {\bf r}_{{\bf i}}\right) =ie\left( \hat{H}_{t{\bf i}%
}^{{\bf a}}-h.c.\right)  \label{b5}
\end{equation}

The first term in Eq. (\ref{b2}) corresponds to a ``paramagnetic''
contribution whereas the second one is of a ``diamagnetic'' origin.

In order to reduce the calculation of physical quantities to a computation
of correlation functions with the AES action we decouple the interaction
term, Eq. (\ref{a3}), by integration over an additional $V_{{\bf i}}\left(
\tau \right) $%
\begin{eqnarray}
\exp \left( -\frac{e^{2}}{2}\sum_{{\bf ij}}\hat{N}_{{\bf i}}C_{{\bf ij}}^{-1}%
\hat{N}_{{\bf j}}\right)  &=&\int \exp \left( -i\sum_{i}\int \left( \psi
^{\ast }\left( {\bf r}_{{\bf i}},\tau \right) \psi \left( {\bf r}_{{\bf i}%
},\tau \right) d{\bf r}_{{\bf i}}-\bar{N}\right) V_{{\bf i}}\left( \tau
\right) d\tau \right)   \nonumber \\
&&\times \exp \left( -\frac{1}{2e^{2}}\sum_{{\bf ij}}\int d\tau V_{{\bf i}%
}\left( \tau \right) C_{{\bf ij}}V_{{\bf j}}\left( \tau \right) \right) DV
\label{c1}
\end{eqnarray}
and then, following Refs. \cite{gefen,belalt}, remove this field from $%
\hat{H}_{0}$ by the gauge transformation 
\begin{equation}
\psi \left( {\bf r}_{{\bf i}}{\bf ,}\tau \right) \rightarrow e^{-i\varphi _{%
{\bf i}}\left( \tau \right) }\psi _{{\bf i}}\left( {\bf r}_{{\bf i}},\tau
\right) ,\text{ \ \ \ \ \ }\dot{\varphi}_{{\bf i}}\left( \tau \right) =V_{%
{\bf i}}\left( \tau \right) .  \label{a4}
\end{equation}
This is not a trivial procedure, because the new fields $\psi _{\alpha
}\left( \tau \right) $ should obey, as before, the fermionic boundary
condition 
\begin{equation}
\psi \left( {\bf r}_{{\bf i}},\tau \right) =-\psi \left( {\bf r}_{{\bf i}%
},\tau +\beta \right) ,\text{ }\beta =1/T  \label{a4a}
\end{equation}
Let us forget for a while about the tunneling between the grains and explain
how one can proceed for the Hamiltonian $H_{0}$ of a single grain.

Due to the boundary condition, Eq. (\ref{a4a}), the field $V_{{\bf i}}\left(
\tau \right) $ cannot be completely removed from $\hat{H}_{0}$ and this
should be done approximately. The field $V_{{\bf i}}\left( \tau \right) $
can be represented as a sum of a static $V_{0{\bf i}}$ part and periodic
function $\tilde{V}_{{\bf i}}\left( \tau \right) $%
\begin{equation}
V_{{\bf i}}\left( \tau \right) =V_{0{\bf i}}+\tilde{V}_{{\bf i}}\left( \tau
\right) ,\text{ \ \ }\int_{0}^{\beta }\tilde{V}_{{\bf i}}\left( \tau \right)
d\tau =0  \label{b6}
\end{equation}
The static part $V_{0}$ can still be arbitrary and we rewrite it as 
\begin{equation}
V_{0{\bf i}}=2\pi Tk_{{\bf i}}+\rho _{{\bf i}},\text{ \ }-\pi T<\rho _{{\bf i%
}}<\pi T\text{\ }  \label{b7}
\end{equation}
where $k_{{\bf i}}=0,\pm 1,$ $\pm 2,....$

If we neglected $\rho _{{\bf i}}$ we would be able to remove $V_{{\bf i}%
}\left( \tau \right) $ from $H_{0}$. In this case we would use, instead of
the phases $\varphi \left( \tau \right) $ from Eq. (\ref{a4}), the phases $%
\tilde{\phi}\left( \tau \right) $ defined as 
\begin{equation}
\tilde{\phi}_{{\bf i}}\left( \tau \right) =\phi _{{\bf i}}\left( \tau
\right) +2\pi Tk_{{\bf i}}\tau ,  \label{a5}
\end{equation}
where $-\infty <\phi _{{\bf i}}\left( \tau \right) <\infty $, $\phi _{{\bf i}%
}\left( 0\right) =\phi _{{\bf i}}\left( \beta \right) .$

It is clear that making the gauge transformation, Eq. (\ref{a4}), with the
phases $\tilde{\phi}_{{\bf i}}\left( \tau \right) $ instead of $\varphi _{%
{\bf i}}(\tau )$, preserves the antiperiodicity of the $\psi \left( \tau
\right) $, Eq. (\ref{a4a}).

The variable $\rho _{{\bf i}}$ cannot generally be neglected. This term is
not important only in the limit $T\gg \delta $. In order to estimate its
contribution we can calculate the partition function $Z^{\left( \rho \right)
}$ of a single grain (normalized by the partition function of the system
without interaction). Carrying out the summation over the Matsubara
frequencies $\varepsilon _{n}$ and making the gauge transformation with the
phases $\tilde{\phi}\left( \tau \right) $, Eq. (\ref{a5}), we obtain 
\begin{equation}
Z^{\left( \rho \right) }=\exp f\left( \rho \right) ,  \label{b8a}
\end{equation}
\[
f\left( \rho \right) =\sum_{\alpha }\left[ \ln \cosh \frac{\xi _{\alpha
}+i\rho }{2T}-\ln \cosh \frac{\xi _{\alpha }}{2T}\right] -i\bar{N}\rho \text{%
,} 
\]
where $\xi _{\alpha }$ is the energy of the state $\alpha $ (the Fermi
energy $\varepsilon _{F}$ is subtracted).

The sum over $\alpha $ in Eq. (\ref{b8a}) extends over all states of the
grain. However, the contribution of the states far from the Fermi energy
must be compensated by the local potentials $\bar{N}.$ As usual, an
essential contribution comes from energies of the order $T$ in the vicinity
of the Fermi energy. The linear in $\rho $ term in the function $f\left(
\rho \right) $, Eq. (\ref{b8a}), must be absent due to the choice of $\bar{N}
$.

A typical separation of the energy levels $\xi _{\alpha }$ is of the order
of the mean level spacing $\delta .$ At not very low temperatures $T\gg
\delta $, the sum over $\alpha $ can be replaced by a proper integral over a
continuous variable $\xi $%
\[
\sum_{\alpha }\rightarrow \frac{1}{\delta }\int d\xi 
\]
and we write the function $f\left( \rho \right) $ in the form 
\begin{equation}
f\left( \rho \right) =\frac{1}{\delta }\int_{-R}^{R}\left[ \ln \cosh \frac{%
\xi +i\rho }{2T}-\ln \cosh \frac{\xi }{2T}\right] d\xi  \label{b9}
\end{equation}
where $R$ is an energy in the range $T\ll R\ll \varepsilon _{F}$. If $\rho $
satisfies the inequality (\ref{b7}) we can deform the contour of integration
when integrating with the first term in Eq. (\ref{b9}) and integrate first
along the straight line ($-R,-R-i\rho $), then along ($-R-i\rho ,$ $R-i\rho $%
) and, at last, along $\left( R-i\rho ,\text{ }R\right) $. The integral over
the second segment cancels the second term in the integrand in Eq. (\ref{b9}%
) and we come to the result 
\begin{equation}
f\left( \rho \right) =-\frac{\rho ^{2}}{2\delta T}  \label{b10}
\end{equation}
Using Eqs. (\ref{b8a}, \ref{b10}) we can understand easily that the main
contribution in integrals over $\rho $ comes from $\rho \sim \left( T\delta
\right) ^{1/2}\ll T$. Therefore, in the limit $T\gg \delta $ the variable $%
\rho $ can safely be put to zero in all calculations and we come to a phase
free energy functional containing only the phases $\tilde{\phi}\left( \tau
\right) ,$ Eq. (\ref{a5}). We see from this argument that introducing the
``winding numbers'' $k_{{\bf i}}$, Eqs. (\ref{a5}, \ref{b7}), is
unavoidable. The opposite limit $T\ll \delta $ is more difficult and is not
considered in the present paper.

Thus, in the limit $T\gg \delta ,$ we are able to remove the effective
voltage $V_{{\bf i}}\left( \tau \right) $ from the single grain Hamiltonian $%
H_{0},$ which means removing the Coulomb interaction. However, the phases
enter now the tunneling Hamiltonian $H_{t}$. Expanding in the tunneling term 
$H_{t}$, Eq. (\ref{a2a}), up to the second order we obtain the AES \cite{aes}
action $S$ in the standard form 
\begin{equation}
S=S_{c}+S_{t},  \label{a9}
\end{equation}
where $S_{c}$ describes the charging energy 
\begin{equation}
S_{c}=\frac{1}{2e^{2}}\sum_{{\bf ij}}\int_{0}^{\beta }d\tau C_{{\bf ij}}%
\frac{d\tilde{\phi}_{{\bf i}}\left( \tau \right) }{d\tau }\frac{d\tilde{\phi}%
_{{\bf j}}\left( \tau \right) }{d\tau }  \label{a10}
\end{equation}
and $S_{t}$ stands for tunneling between the grains 
\begin{equation}
S_{t}=\pi g\sum_{\left| {\bf i}-{\bf j}\right| =a}\int_{0}^{\beta }d\tau
d\tau ^{\prime }\alpha \left( \tau -\tau ^{\prime }\right) \sin ^{2}\left( 
\frac{\tilde{\phi}_{{\bf ij}}\left( \tau \right) -\tilde{\phi}_{{\bf ij}%
}\left( \tau ^{\prime }\right) }{2}\right)  \label{a11}
\end{equation}

The function $\alpha \left( \tau \right) $ in Eq. (\ref{a11}) has the form 
\[
\alpha \left( \tau \right) =T^{2}\left( 
\mathop{\rm Re}%
\left( \sin \left( \pi T\tau +i\delta \right) \right) ^{-1}\right) ^{2}. 
\]
In Eqs. (\ref{a9}-\ref{a11}), ${\bf i}$ and ${\bf j}$ stand for coordinates
of grains, $a$ is the diameter of a grain. The dimensionless conductance $g$
is given by 
\begin{equation}
g=2\pi \nu _{0}^{2}t_{{\bf ij}}^{2}  \label{b11}
\end{equation}
where $t_{{\bf ij}}$ is the tunneling amplitude from grain ${\bf i}$ to
grain ${\bf j}$ (spin is included) and $\nu _{0}$ is the density of states
of non-interacting electrons.

Calculation of different averages with the functional $S,$ Eqs. (\ref{a9}-%
\ref{a11}), implies integration over $\phi _{{\bf i}}\left( \tau \right) $
and summation over $k_{{\bf i}}$. At large $g\gg 1$, one can put all $k_{%
{\bf i}}=0$. However, at $\ g\lesssim 1$ one should sum over all $k_{{\bf i}}
$ and neglecting the contribution of the non-zero winding numbers leads to
incorrect results. We emphasize that the AES free energy functional can be
derived under the assumption that the temperatures are not very low, Eqs. (%
\ref{ad1}, \ref{ad2}), and is applicable in this limit only. At lower
temperatures one would have to take into account interference effects \cite
{belalt} or the discreteness of the levels in single grains.

Using the Kubo formulae, Eqs. (\ref{b1}- \ref{b5}), we can express also the
conductivity in terms of the phases $\tilde{\phi}_{{\bf i}}\left( \tau
\right) $. After simple manipulations we represent the conductivity $\sigma
\left( \omega \right) $ as

\begin{equation}
\sigma \left( \omega \right) =\frac{ia^{2-d}}{\omega }\left[ \int_{0}^{\beta
}d\tau e^{i\Omega _{n}\tau }K\left( \tau \right) \right] _{\Omega
_{n}\rightarrow -i\omega +\delta },  \label{a7}
\end{equation}

\[
K\left( \tau \right) =\langle X_{2}^{{\bf a}}\left( \tau \right) \rangle
-\sum_{{\bf i}}\langle X_{10}^{{\bf a}}\left( \tau \right) X_{1{\bf i}}^{%
{\bf a}}\left( 0\right) \rangle , 
\]
\[
X_{2}^{{\bf a}}\left( \tau \right) =e^{2}\pi g\int_{0}^{\beta }d\tau
^{\prime }\left( \delta \left( \tau \right) -\delta \left( \tau ^{\prime
}-\tau \right) \right) \alpha \left( \tau ^{\prime }\right) \cos \left( 
\tilde{\phi}_{{\bf i,i+a}}\left( \tau ^{\prime }\right) -\tilde{\phi}_{{\bf %
i,i+a}}\left( 0\right) \right) , 
\]
\[
X_{1{\bf i}}^{{\bf a}}\left( \tau \right) =e\pi g\int_{0}^{\beta }\alpha
\left( \tau -\tau ^{\prime }\right) \sin \left( \tilde{\phi}_{{\bf i,i+a}%
}\left( \tau ^{\prime }\right) -\tilde{\phi}_{{\bf i,i+a}}\left( \tau
\right) \right) d\tau ^{\prime }, 
\]

In Eqs. (\ref{a7}), $\tilde{\phi}_{{\bf ij}}\left( \tau \right) =\tilde{\phi}%
_{{\bf i}}\left( \tau \right) -\tilde{\phi}_{{\bf j}}\left( \tau \right) $
for ${\bf i}$ and ${\bf j}$ standing for neighboring grains and 
\begin{equation}
\langle ...\rangle =\int \left( ...\right) \exp \left( -S\right) D\tilde{\phi%
}\left( \int \exp \left( -S\right) D\tilde{\phi}\right) ^{-1},  \label{a8}
\end{equation}
where $D\tilde{\phi}$ implies both the functional integration over $\phi
\left( \tau \right) $ and summation over the winding numbers $k_{{\bf i}}$.

Eqs. (\ref{a9}-\ref{a8}) represent the conductivity $\sigma \left( \omega
\right) $ in a closed form in terms of a functional integral. The
contribution of the function $X_{2}\left( \tau \right) $ originates from the
diamagnetic term (the second term in Eq. (\ref{b2})), whereas the
correlation function $<X_{1}X_{1}>$ comes from the paramagnetic term $\Pi ,$
Eq. (\ref{b3}) (the first term in Eq. (\ref{b2})).

Although the model described by Eqs. (\ref{a9}-\ref{a8}) is simpler than the
initial model, Eqs. (\ref{a2}-\ref{a3}), explicit formulae can be written
only in limiting cases.

If the temperature $T$ is very high, $T\gg E_{c}\sim e^{2}C_{ij}^{-1}$,
where $E_{c}$ is the electrostatic energy of adding one electron to a grain,
fluctuations of the phases $\tilde{\phi}$ are negligible and one can set $%
\tilde{\phi}=0$ in the expressions for $X_{1\text{ }}$and $X_{2}$ in Eqs. (%
\ref{a7}). Then, we obtain easily the conductivity 
\begin{equation}
\sigma _{0}=e^{2}ga^{2-d}  \label{a12}
\end{equation}
Eq. (\ref{a12}) describes the classical conductivity of the granular metal
without the Coulomb interaction and shows that at temperatures exceeding $%
E_{c}$ charging effects are not important.

In the opposite limit $T\ll E_{c},$ transport in the granulated system has
much more interesting characteristics. This inequality can be compatible
with the inequality $T\gg \max \{g\delta ,\delta \},$ used for the
derivation of Eqs. (\ref{a9}-\ref{a8}), because $E_{c}\gg \delta $ for $2D$
and $3D$ grains (the charging energy $E_{c}$ is inversely proportional to
size of the grain, while $\delta $ is inversely proportional to its volume).

At temperatures $T\lesssim E_{c}$, calculations are possible only in the
limiting cases $g\gg 1$ and $g\ll 1$ and this will be done in the subsequent
sections.

The same action $S$, Eqs. (\ref{a9}-\ref{a11}), was used in Ref. \cite{fazio}%
, and a metal-insulator transition has been predicted in a $2D$ array of
tunnel junctions. However, the authors of Ref. \cite{fazio} did not
calculate the conductivity but discussed properties of the partition
function. For large $g$ they did not account for phase fluctuations properly
which, as we show here, are responsible for the behavior, Eq. (\ref{a1a}).
Moreover, we find a transition in any dimensionality.

\section{Conductivity in the metallic regime at $g\gg 1$}

\subsection{Perturbation theory}

In the limit of large conductances $g\gg 1$, the tunnelling term, Eq. (\ref
{a11}), suppresses large fluctuations of $\phi $. It is clear that all
non-zero winding numbers $k_{{\bf i}}$ can be neglected. Accounting for
non-zero $k_{{\bf i}}$ (as well as variations of $\bar{N}_{{\bf i}}$ ) would
lead to contributions of order $\exp \left( -g\right) ,$ which can be
neglected in any expansion in $1/g$. At the same time, the phase
fluctuations can change considerably the classical result, Eq. (\ref{a12}),
even in the limit $g\gg 1$.

Let us understand first the role of the fluctuations within a perturbation
theory in $1/g$.

The zeroth order of the perturbation theory (all phases $\tilde{\phi}$ are
set to zero), gives for the conductivity the classical result $\sigma _{0}$,
Eq. (\ref{a12}). In order to consider higher orders we expand the action $S$%
, Eqs. (\ref{a9}-\ref{a11}), in $\phi $.

The quadratic part $S_{2}$ of the action $S$ will serve as the bare action
in the perturbation theory we want to develope now. Keeping terms of the
second order in $\phi $ in Eqs. (\ref{a9}-\ref{a11}) and performing Fourier
transformation in both coordinates of the grains and the imaginary time we
reduce the action $S$ to the form 
\begin{equation}
S_{0}=T\sum_{{\bf q},n}\phi _{{\bf q},n}G_{{\bf q},n}^{-1}\phi _{-{\bf q}%
,-n},  \label{a13}
\end{equation}
\begin{equation}
G_{{\bf q},n}^{-1}=\omega _{n}^{2}/\left( 4E\left( {\bf q}\right) \right)
+2g\left| \omega _{n}\right| \sum_{{\bf a}}\left( 1-\cos {\bf qa}\right) ,
\label{a13a}
\end{equation}
where $E\left( {\bf q}\right) =e^{2}/\left( 2C\left( {\bf q}\right) \right) $
and $C\left( {\bf q}\right) $ is the Fourier-transform of the capacitance
matrix $C_{{\bf ij}}$ (${\bf q}$ are quasi-momenta for the array of the
grains). One should sum in Eq. (\ref{a13}) over $d$ unit lattice vectors $%
{\bf a,}$ where $d$ is the dimensionality of the array.

If we kept only quadratic in $\phi $ terms in the action $S$ but did not
expand the function $X_{2}$, Eq. (\ref{a7}), we would reduce the correlation
function $\langle X_{2{\bf a}}\left( \Omega _{n}\right) \rangle $ to the
form 
\begin{equation}
\langle X_{2}^{{\bf a}}\left( \Omega _{n}\right) \rangle _{0}=\pi
e^{2}g\int_{0}^{\beta }\alpha \left( \tau \right) \left( 1-e^{i\Omega
_{n}\tau }\right) e^{-\tilde{G_{{\bf a}}}\left( \tau \right) }d\tau ,
\label{a14a}
\end{equation}

\begin{equation}
\tilde{G}_{{\bf a}}\left( \tau \right) =4Ta^{d}\sum_{\omega _{n}>0}\int 
\frac{d{\bf q}}{\left( 2\pi \right) ^{d}}G_{{\bf q}n}\sin ^{2}\frac{{\bf qa}%
}{2}\sin ^{2}\frac{\omega _{n}\tau }{2}  \label{a14}
\end{equation}
where $<...>_{0}$ means averaging over the phases $\phi $ with the action $%
S_{0}$, Eq. (\ref{a13}) and $\omega _{n}=2\pi n$. What remains to be done in
order to calculate the contribution $X_{2}^{\left( 0\right) }$ to the
conductivity is to compute the integral in Eq. (\ref{a14a}) for the
Matsubara frequencies $\Omega _{n}$ and make the analytical continuation $%
\Omega _{n}\rightarrow -i\omega +\delta $. As it is clear from Eqs. (\ref
{a13a}, \ref{a14}), the function $\tilde{G}_{{\bf a}}\left( \tau \right) $
contains large logarithms $\ln \left( gE_{c}\tau \right) $ and essential $%
\tau $ are of the order $1/T$. Therefore, we may calculate the integral for $%
\tilde{G}_{{\bf a}}\left( \tau \right) $ with a logarithmic accuracy.
Neglecting the $\omega _{n}^{2}$ term in $G_{{\bf q,}n}^{-1}$, Eq. (\ref
{a13a}), we reduce Eq. (\ref{a14}) to the form 
\begin{equation}
\tilde{G}_{{\bf a}}\left( \tau \right) =\frac{T}{dg}\sum_{\omega
_{n}>0}^{\omega _{c}}\frac{1-\cos \left( \omega _{n}\tau \right) }{\omega
_{n}}  \label{b12}
\end{equation}
In Eq. (\ref{b12}) one should sum over positive Matsubara frequencies up to
the cutoff $\omega _{c}\sim gE_{c}$. Eq. (\ref{b12}) shows a remarkable
independence of the result on the structure of the lattice. The only
information about the lattice is the parameter $d$ entering Eq. (\ref{b12}).
For the cubic lattice considered here, $d$ is the dimensionality of the
array. However, for an arbitrary lattice the parameter $d$ is equal to the
one half of the coordination number. What is also important, there are no
``infrared'' divergencies in the integral over ${\bf q}$ in any
dimensionality including $2D$ and $1D$. This is specific for the
conductivity. We will see later that the tunnelling density of states $\nu $
is sensitive to the dimensionality in the same approximation due to infrared
divergencies.

As we are performing the calculations with the logarithmic accuracy, we may
replace $\tau $ by $1/T$ in the function $\tilde{G}_{{\bf a}}$ and calculate
the remaining integral over $\tau $ in Eq. (\ref{a14a}) ignoring the
dependence of the function $\tilde{G}_{{\bf a}}$ on $\tau $. Then, we obtain 
\begin{equation}
<X_{2}\left( \omega \right) >_{0}=-i\omega e^{2}g\left( \frac{T}{gE_{c}}%
\right) ^{\alpha }  \label{b13}
\end{equation}
where 
\begin{equation}
\text{\ }\alpha =\left( 2\pi gd\right) ^{-1}  \label{b13a}
\end{equation}

This form of the correlation function $<X_{2}\left( \omega \right) >$ would
lead to the power law dependence of the conductivity on temperature of the
form of Eq. (\ref{a1}). A similar result has been obtained for the voltage
dependence of the conductance of a single junction in a model with an
electromagnetic environment \cite{devoret,girvin}. If Eq. (\ref{a1}) were
the final result of our calculations we could argue that, in the model under
consideration, fluctuations of the phases and, hence, of the voltages in the
grains are equivalent to fluctuations in the electromagnetic environment of
the works \cite{devoret,girvin}. In other words, each grain would be
considered as surrounded by an effective medium of other grains and the
voltage fluctuations of the medium would lead to the power law, Eq. (\ref{a1}%
).

However, Eq. (\ref{b13}), is not the final result yet because we have to
calculate also the contribution coming from higher order terms of the
expansion of the action $S$, Eq. (\ref{a9}-\ref{a11}), in $\phi $ as well as
the contribution of the correlation function $<X_{1}X_{1}>$, Eq. (\ref{a7}).
Taking into account these contributions can be performed writing an
expansion for the conductivity in powers of $1/g$. The first three terms $%
\sigma ^{\left( 1\right) }$ of the expansion of the conductivity in $1/g$
coming from the function $<X_{2}\left( \omega \right) >_{0}$, (\ref{b13})
can be written as 
\begin{equation}
\sigma ^{\left( 1\right) }/\sigma _{0}=1-\alpha \ln \left( \frac{gE_{c}}{T}%
\right) +\frac{\alpha ^{2}}{2}\ln ^{2}\left( \frac{gE_{c}}{T}\right)
\label{b14}
\end{equation}
and we want to find now contributions of the order up to $\alpha ^{2}$
coming from $<X_{1}X_{1}>$ and those originating from $S_{4}$, where $S_{4}$
contains terms of order $\phi ^{4}$ in the action $S$, Eqs. (\ref{a9}-\ref
{a11}).

As concerns a contribution coming from the correlation function $%
<X_{1}X_{1}> $, the first non-vanishing term is of the order $\alpha ^{2}$
and it does not contain powers of $\ln \left( gE_{c}/T\right) $. So, we
neglect the function $<X_{1}X_{1}>$ and concentrate on the contribution to
the function $<X_{2}\left( \omega \right) >$ coming from the anharmonic part 
$S_{4}$ of the action $S$. It is clear from a power counting that higher
order terms of the expansion of $S$ lead to contributions containing higher
powers of $\alpha $ and we do not consider them now.

The lowest order contribution coming from $S_{4}$ is obtained by averaging
with the action $S_{0}$, Eq. (\ref{a13}), of a product of a term $\phi ^{2}$
taken from the expansion of $X_{2}$, Eq. (\ref{a7}), and $S_{4}$. A proper
expression can be written as 
\begin{equation}
<\sum_{{\bf i}_{1},{\bf a}_{1}}\int_{0}^{\beta }\int_{0}^{\beta
}\int_{0}^{\beta }d\tau d\tau _{1}d\tau _{1}^{\prime }{}\left( e^{i\Omega
_{n}\tau }-1\right) \alpha \left( \tau \right) \alpha \left( \tau _{1}-\tau
_{1}^{\prime }\right)  \label{b15}
\end{equation}
\[
\times \left( \phi _{{\bf i+a}_{0},{\bf i}}\left( \tau \right) -\phi _{{\bf %
i+a}_{0},{\bf i}}\left( 0\right) \right) ^{2}\left( \phi _{{\bf i}_{1}+{\bf %
a,i}_{1}}\left( \tau _{1}\right) -\phi _{{\bf i}_{1}+{\bf a,i}_{1}}\left(
\tau _{1}^{\prime }\right) \right) ^{4}>_{0} 
\]
After Fourier transforming the phases in both coordinates and time we can
average easily with the action $S_{0}$, Eq. (\ref{a13}). Then, Eq. (\ref{b15}%
) is reduced to the form 
\[
12T^{3}\sum_{{\bf a,\omega ,q}_{1},{\bf q}_{2},{\bf q}_{3}}G_{{\bf q}%
_{1},n_{1}}^{2}G_{{\bf q}_{2},n_{2}}|e^{i{\bf q}_{1}{\bf a}%
_{0}}-1|^{2}\prod_{i=1,2}|e^{i{\bf q}_{i}{\bf a}}-1|^{2} 
\]
\begin{equation}
\times \int_{0}^{\beta }\left( e^{i\Omega _{n}\tau }-1\right) |e^{-i\omega
_{n1}\tau }-1|^{2}\alpha \left( \tau \right) d\tau \int_{0}^{\beta
}|1-e^{i\omega _{n1}\tau }|^{2}|1-e^{i\omega _{n2}\tau }|^{2}\alpha \left(
\tau \right) d\tau  \label{b16}
\end{equation}
Integrals over $\tau $ can easily be calculated by changing to the variables 
$z=\exp \left( 2\pi iT\tau \right) $. The lattice integrations are trivial
as well and we come to the contribution to the conductivity $\sigma ^{\left(
2\right) }$coming from Eq. (\ref{b15}) 
\begin{equation}
\sigma ^{\left( 2\right) }/\sigma _{0}=-\alpha ^{2}\sum_{n_{1}>n_{2}>0}\frac{%
1}{n_{1}n_{2}}=-\frac{\alpha ^{2}}{2}\ln ^{2}\left( \frac{gE_{c}}{T}\right)
\label{b17}
\end{equation}
Adding the contributions $\sigma ^{\left( 1\right) }$ and $\sigma ^{\left(
2\right) }$ we obtain for the conductivity $\sigma $ in the limit $\omega
\rightarrow 0$%
\begin{equation}
\sigma =\sigma _{0}\left( 1-\alpha \ln \left( \frac{gE_{c}}{T}\right) \right)
\label{b18}
\end{equation}
We see that the terms of the order $\alpha ^{2}\ln ^{2}\left(
gE_{c}/T\right) $ in $\sigma ^{\left( 1\right) }$ and $\sigma ^{\left(
2\right) }$ cancel each other and the accuracy of Eq. (\ref{b18}) exceeds $%
\alpha \ln \left( gE_{c}/T\right) $. We emphasize again that, in this
approximation, there is no dependence on the structure of the lattice of the
grains except that the dimensionality \ (one half of the coordination
number) $d$ enters $\alpha $, Eq. (\ref{b13a}). However, this property holds
only for contributions of the type $\alpha ^{n}\ln ^{n}\left(
gE_{c}/T\right) $. Terms with lower powers of logarithms depend on the
structure of the lattice in a more complicated way.

The cancellation of the terms of the order $\alpha ^{2}\ln ^{2}\left(
gE_{c}/T\right) $ when calculating $\sigma $, Eq. (\ref{b18}), is not
accidental and we want to demonstrate this within a renormalization group
(RG) scheme.

\subsection{Renormalization Group}

In order to sum up the logarithmic corrections to the conductivity we use RG
arguments suggested for a one-dimensional inverse square $XY$-model long ago 
\cite{kosterlitz} and used later in a number of works \cite
{schmid,bulg,guinea,falci,zwerger}. We assume that the tunnel conductance is
large, $g\gg 1$ and therefore we use the phases $\phi $ neglecting the
winding numbers. As the starting functional we take the tunneling action $%
S_{t}$%
\begin{equation}
S_{t}=\pi g\sum_{\left| {\bf i}-{\bf j}\right| =a}\int_{0}^{\beta
}\int_{0}^{\beta }d\tau d\tau ^{\prime }\alpha \left( \tau -\tau ^{\prime
}\right) \sin ^{2}\left( \frac{\phi _{{\bf ij}}\left( \tau \right) -\phi _{%
{\bf ij}}\left( \tau ^{\prime }\right) }{2}\right)  \label{b19}
\end{equation}
This action contains the conductance $g$, which determines the conductivity $%
\sigma $. The charging part $S_{c}$ is not important for the renormalization
group because it determines only the upper cutoff of integrations over
frequencies. In the limit $T\rightarrow 0$ the function $\alpha $ is
proportional to $\left( \tau -\tau ^{\prime }\right) ^{-2}$ and the action
is dimensionless.

Following standard RG arguments we want to find how the form of the action $%
S_{t}$ changes under changing cutoffs. Generally speaking, it is not
guaranteed that after integrating over the phases $\phi $ in an interval of
frequencies one comes to the same function $\sin ^{2}\phi $ in the action.
The form of the functional may change, which would lead to a functional
renormalization group. In the present case appearance of terms $\sin
^{2}2\phi $, $\sin ^{2}4\phi $, etc. is not excluded and, indeed, they are
generated in many loop approximations of the RG. Fortunately, the one loop
approximation is simpler and the renormalization in this order results in a
change of the effective coupling constant $g$ only.

To derive the RG equation we represent the phase $\phi $ in the form 
\begin{equation}
\phi _{{\bf ij\omega }}=\phi _{{\bf ij\omega }}^{\left( 0\right) }+\overline{%
\phi }_{{\bf ij\omega }}  \label{b20}
\end{equation}
The function $\phi _{{\bf ij\omega }}$ is not equal to zero in an interval
of the frequencies $0<\omega <\omega _{c}$, while the function $\phi _{{\bf %
ij\omega }}^{\left( 0\right) }$ is finite in the interval $\ \lambda \omega
_{c}<\omega <\omega _{c}$ where $\lambda $ is in the interval $0<\lambda <1$%
. Integrating in the expression for the partition function 
\[
Z=\int \exp \left( -S_{t}\right) D\phi 
\]
over the function $\phi _{{\bf ij\omega }}^{\left( 0\right) }$ \ we come to
a new action $\overline{S}$ with the cutoff $\lambda \omega _{c}$.
Substituting Eq. (\ref{b20}) into Eq. (\ref{b19}) we expand the action $%
S_{t} $ up to terms quadratic in $\phi _{{\bf ij}\omega }^{\left( 0\right) }$%
. Integrating over $\phi _{{\bf ij\omega }}^{\left( 0\right) }$ is
straightforward and we obtain with the logarithmic accuracy a renormalized
effective action $\tilde{S}_{t}$%
\begin{equation}
\tilde{S}_{t}=2\pi g\sum_{\left| {\bf i}-{\bf j}\right| =1}\int_{0}^{\beta
}\int_{0}^{\beta }d\tau d\tau ^{\prime }\alpha \left( \tau -\tau ^{\prime
}\right) \sin ^{2}\left( \frac{\phi _{{\bf ij}}\left( \tau \right) -\phi _{%
{\bf ij}}\left( \tau ^{\prime }\right) }{2}\right) \left( 1-\frac{\xi }{2\pi
gd}\right)  \label{b21}
\end{equation}
where $\xi =-\ln \lambda $.

We see from Eq. (\ref{b21}) that the form of the action is reproduced for
any dimensionality $d$ of the lattice of the grains. This allows us to write
immediately the following renormalization group equation 
\begin{equation}
\frac{\partial g\left( \xi \right) }{\partial \xi }=-\frac{1}{2\pi d}
\label{b22}
\end{equation}
The solution of Eq. (\ref{b22}) is simple. Neglecting the Coulomb
interaction in the action $S_{t}$, Eq. (\ref{b19}), is justified only for
energies smaller than $gE_{c}$ and this energy is the upper cutoff. Then the
renormalized conductance $g\left( T\right) $ takes the form 
\begin{equation}
g\left( T\right) =g-\frac{1}{2\pi d}\ln \frac{gE_{c}}{T}  \label{b23}
\end{equation}
and we come to Eq. (\ref{b18}) for conductivity. Both the quantities depend
on the temperature logarithmically.

Eqs. (\ref{b23}) is obtained in the one loop approximation and should be
valid so long as the effective conductance $g\left( T\right) $ remains much
larger than $1$. Therefore, Eq. (\ref{b18}) is also more than a result of
the perturbation theory and is valid for $\sigma /\sigma _{0}\gg 1/g$. This
gives the necessary condition for the applicability of Eqs. (\ref{b18}, \ref
{b23}) 
\begin{equation}
1-\left( 2\pi dg\right) ^{-1}\ln \left( gE_{c}/T\right) \gg 1/g  \label{b24}
\end{equation}

However, the condition (\ref{b24}) is not sufficient for the applicability
of Eqs. (\ref{b18},\ref{b23}) because the AES action may be used only in the
limit 
\begin{equation}
T\gg g\delta  \label{b25}
\end{equation}
\bigskip If the conductance $g$ or the size of the grains are not very large
the condition (\ref{b25}) can become stronger than (\ref{b24}). Then, at
lower temperatures one should take into account interference effects and,
depending on the dimensionality $d$ of the array, both metal and insulating
states are possible. In contrast, Eqs. (\ref{b18},\ref{b23}), are valid in 
{\em any }dimensionality.

Actually, in the main approximation the action $S_{t}$ contains only the
phase differences $\phi _{{\bf ij}}\left( \tau \right) $ that can be
considered as variables on sites of a lattice dual to the original one.
These sites are not coupled in this approximation with each other and this
explains why we obtained the same results as those for one contact. Taking
into account the charging energy $S_{c}$, Eq. (\ref{a10}), couples the sites
and the structure of the lattice may become important in next orders of the
renormalization group equations.

One can check that the contribution coming from the correlator of the
functions $X_{1}$ in Eqs. (\ref{a7}) contains additional powers of $1/g$ and
can be neglected in the main approximation. It is very important that the
correlator $\langle X_{1}X_{1}\rangle $ in Eqs. (\ref{a7}) representing the
``paramagnetic contribution'' contains a summation over $j,$ which
corresponds to the zero quasi-momentum of the function $K$. Keeping in $%
X_{1} $ linear terms in $\phi $ would give zero even if anharmonic terms in $%
S$ would have been taken into account. If we carried out the computation for
a single grain the contribution from the $\langle X_{1}X_{1}\rangle $ would
not be smaller than that coming from $\langle X_{2}\rangle $.

It is very important to note that we use the linear response theory for the
calculation of the conductivity assuming that the external electric field is
homogeneous. This contrasts calculation of the conductance for a single
contact. In our calculations we do not obtain a contribution to the
conductivity corresponding to the inelastic co-tunnelling known for single
dots \cite{nazarov}. This is natural because in the problem considered the
inelastic co-tunnelling can occur only through the entire system. This would
lead to contributions exponentially small in the size of the system.

\section{Tunneling density of states at $g\gg 1$}

The tunneling density of states (DOS) $\nu _{{\bf i}}\left( \varepsilon
\right) $ in the grain ${\bf i}$ can be introduced in a standard way through
the retarded single particle Green function $G_{{\bf ii}}^{R}\left(
\varepsilon \right) $ with both the coordinates in the grain ${\bf i}$. As
we use here the imaginary time representation we calculate first the
temperature Green function ${\cal G}_{{\bf ii}}\left( \tau \right) $ at
Matsubara frequencies $\varepsilon _{n}=\pi T\left( 2n+1\right) .$ This
leads to a function $\tilde{\nu}_{{\bf i}}\left( \varepsilon _{n}\right) $ 
\begin{equation}
\tilde{\nu}_{{\bf i}}\left( \varepsilon _{n}\right) =-\pi ^{-1}\int d\tau
e^{i\varepsilon _{n}\tau }{\cal G}_{{\bf ii}}\left( \tau \right)  \label{c1a}
\end{equation}
The DOS $\nu _{{\bf i}}\left( \varepsilon \right) $ can be found by the
analytical continuation 
\begin{equation}
\nu _{{\bf i}}\left( \varepsilon \right) =%
\mathop{\rm Im}%
\left( \left. \tilde{\nu}_{{\bf i}}\left( \varepsilon _{n}\right) \right|
_{\varepsilon _{n}\rightarrow -i\varepsilon +\delta }\right)  \label{c2}
\end{equation}
Following the same procedure as the one used in the previous sections we
perform the gauge transformation, Eq. (\ref{a4}), reducing the calculation
to integration over the phases $\phi _{{\bf i}}\left( \tau \right) $. As a
result, we obtain for the function $\tilde{\nu}_{i}\left( \varepsilon
_{n}\right) $%
\begin{equation}
\tilde{\nu}_{{\bf i}}\left( \varepsilon _{n}\right) =\nu
_{0}T\int_{0}^{\beta }d\tau \frac{e^{i\varepsilon _{n}\tau }}{\sin \pi T\tau 
}\left\langle \exp \left( -i\left( \tilde{\phi}_{{\bf i}}\left( \tau \right)
-\tilde{\phi}_{{\bf i}}\left( 0\right) \right) \right) \right\rangle
\label{c3}
\end{equation}
where the symbol $\left\langle ...\right\rangle $ means as before averaging
with the action $S$, Eqs. (\ref{a9}-\ref{a11}) and integration over $\tilde{%
\phi}_{{\bf i}}\left( \tau \right) $ includes summation over the winding
numbers $k_{i}$. The principal value of the integral over $\tau $ is implied
in Eq. (\ref{c3}).

Calculation of DOS is quite different from that for the conductivity because
Eq. (\ref{c3}) contains the phases $\tilde{\phi}_{{\bf i}}\left( \tau
\right) $ corresponding to the grain ${\bf i}$ but not the phase differences 
$\tilde{\phi}_{{\bf ij}}\left( \tau \right) $. This can lead to a
non-trivial dependence of the final result on the dimensionality $d$ of the
lattice.

In the limit of large $g\gg 1$ we expand, as when calculating the
conductivity, the functional $S$, Eqs. (\ref{a9}-\ref{a11}), in $\phi _{{\bf %
ij}}\left( \tau \right) $. If the $d=3$ one should expand in \ $\phi \left(
\tau \right) $ both the action $S$ and the exponential in Eq. (\ref{c3}).
The reason is the same as for calculation of the conductivity: all terms of
the expansion give additional logarithms and there is no reason to keep in
the action only quadratic terms and at the same time not to expand the
exponential in Eq. (\ref{c3}). Therefore in the $3D$ case one can expect
logarithmic corrections to the tunneling density of states with coefficients
depending on the structure of the lattice. Making analytical continuation
onto real energies $\varepsilon $ we can write the tunneling density of
states $\nu _{3}\left( \varepsilon \right) $ in $3D$ in the main
approximation as 
\begin{equation}
\nu _{3}\left( \varepsilon \right) =\nu _{0}T\int_{0}^{\beta }\frac{%
e^{i\varepsilon _{n}\tau }}{\sin \pi T\tau }\left( 1-G\left( \tau \right)
\right) d\tau  \label{c4}
\end{equation}
where 
\begin{equation}
G\left( \tau \right) =2Ta^{d}\sum_{\omega _{n}>0}\int \frac{d^{d}{\bf q}}{%
\left( 2\pi \right) ^{d}}G_{{\bf q}n}\sin ^{2}\frac{\omega _{n}\tau }{2}
\label{c4a}
\end{equation}

and $G_{{\bf q}n}$ is given by Eq. (\ref{a13a}).

Calculating the sum over $\omega _{n}$ with the logarithmic accuracy and
making the analytical continuation we obtain 
\begin{equation}
\nu _{3}\left( \varepsilon \right) =\nu _{0}\left( 1-\frac{A}{4\pi g}\ln
\left( \frac{gE_{c}}{\max \left( T,\varepsilon \right) }\right) \right)
\label{c5}
\end{equation}
{\bf \ } where 
\begin{equation}
A=\int \frac{a^{3}d^{3}{\bf q}}{\left( 2\pi \right) ^{3}}\frac{1}{\sum_{{\bf %
a}}\left( 1-\cos {\bf qa}\right) }  \label{c6}
\end{equation}

Eq. (\ref{c5}) shows that the density of states $\nu _{3}\left( \varepsilon
\right) $ of the $3D$ array of the grains has a logarithmic dependence on
temperature like the conductivity $\sigma $, Eq. (\ref{b18}). However, in
contrast to the latter, Eq. (\ref{c6}) is valid only if the logarithmic term
is much smaller than $1$.

The situation is more interesting in one and two dimensional systems. In
this cases the integral over ${\bf q}$ in Eq. (\ref{c6}) formally diverges,
which means that one should take into account the $\omega ^{2}$ term in the
function $G_{{\bf q}n}$, Eq. (\ref{a13a}). This term cuts the infrared
divergency in the integral over ${\bf q}$ but one obtains in the density of
states a stronger singularity in $\varepsilon ,T$ than the one in $3D$, Eq. (%
\ref{c5}).

Fortunately, this makes the calculation even easier and allows us to obtain
explicit results in the non-perturbative regime when the DOS considerably
deviates from $\nu _{0}$. This simplification is due to the fact that the
strongest singularities come from the expansion in $\phi _{{\bf i}}\left(
\tau \right) $ of the exponential in Eq. (\ref{c3}). The anomalous
contributions originating from small $q$ arise in the expansion of $\phi _{%
{\bf i}}$ but cancel each other in the contributions coming from expansions
in $\phi _{{\bf ij}}\left( \tau \right) =\phi _{{\bf i}}\left( \tau \right)
-\phi _{{\bf j}}\left( \tau \right) $ in the action $S_{t}$, Eq. (\ref{a11}%
). Therefore, expanding in $\phi _{{\bf ij}}\left( \tau \right) $ one
obtains usual logarithmic contributions of the type $\left( g\ln \left(
gE_{c}/T\right) \right) ^{n},$ which allows us to keep in the action $S$
only quadratic in $\phi _{{\bf ij}}\left( \tau \right) $ terms.

Then the integration over $\phi \left( \tau \right) $ in Eq. (\ref{c3}) can
easily be carried out and we obtain for the low dimensions 
\begin{equation}
\nu \left( \varepsilon \right) =\nu _{0}T\int_{0}^{\beta }\frac{%
e^{i\varepsilon _{n}\tau }}{\sin \pi T\tau }\exp \left( -G\left( \tau
\right) \right) d\tau  \label{c7}
\end{equation}
It is clear that the singularity in the exponent in Eq. (\ref{c7}) is
stronger than logarithmic and this justifies the approximation used in the
derivation.

In the most interesting case of a $2D$ array of the granules both the
summation over the frequencies and integration over the momenta ${\bf q}$
give logarithms and we come to the final result 
\begin{equation}
\nu _{2}\left( \varepsilon \right) =\nu _{0}\exp \left( -\frac{1}{16\pi ^{2}g%
}\ln ^{2}\left( \frac{gE_{c}}{\max \left( \varepsilon ,T\right) }\right)
\right)  \label{c8}
\end{equation}
Eq. (\ref{c8}) is valid down to $\max \left( \varepsilon ,T\right) \sim
g\delta $ when the description in terms of the phase functional $S$, Eqs. (%
\ref{a9}-\ref{a11}), is still applicable. Eq. (\ref{c8}) perfectly agrees
with the corresponding result obtained long ago for disordered films using a
replica $\sigma $-model \cite{fin}. This result was reproduced for
disordered films in a number of subsequent publications \cite
{naz,levit,kop,andreev} using different approaches. The strong anomaly in
the exponent in Eq. (\ref{c8}) is due to the fact that the one particle
Green function is not gauge invariant\cite{fin}. The singularity is formed
by almost pure gauge fluctuations of the electric fields. Gauge-invariant
characteristics like conductivity are not influenced by such fluctuations
and therefore are less anomalous.

The applicability of Eq. (\ref{c8}) not only for disordered films but also
for the granular systems at not very low temperatures shows that the result
is very robust. In contrast, the dependence of the conductivity on
temperature, Eqs. (\ref{b18}-\ref{b23}), cannot be used for very low
temperatures or disordered ``homogeneous systems''. The formal reason for
this difference is quite clear: the main contribution to the conductivity
comes from momenta $q\sim a^{-1}$, whereas the main contribution to the
density of states comes from small $q\ll a^{-1}$. The latter limit is not
sensitive to the structure of the system at short distances.

The coefficient in front of $\ln ^{2}$ is somewhat different than that of
the recent work \cite{andreev}. This is because we assumed that $E\left( 
{\bf q}\right) $ remains finite in the limit of $q\rightarrow 0$ (there is a
screening in the system). In the integral over ${\bf q}$ in Eq. (\ref{c4a}, 
\ref{a13a}), essential $\left( qa\right) ^{2}$ were of the order of $\omega
/gE_{c}$. In contrast, if one starts with a non-screened $2D$ Coulomb
interaction $V_{0}=2\pi e^{2}/q$, the essential $q$ are proportional to $%
\omega $ and this increases the coefficient by the factor of $2$.

The first term of the expansion of the exponential in Eq. (\ref{c8}) is just
the Altshuler-Aronov correction generalized to the case of the granular
metal \cite{altar}. Of course, the same is true for the correction to the
conductivity, Eq. (\ref{b18}). This is quite natural because at not very low
temperatures weak localization corrections are suppressed but the
Altshuler-Aronov corrections still give important contributions. Actually,
the function $|\omega _{n}|G_{{\bf q}}$, Eq.(\ref{a13a}), is just the
Coulomb propagator screened by the electron-electron interaction. In other
words, the theory developed in the previous sections for $g\gg 1$ starting
from the AES action is another way of calculation of the Altshuler-Aronov
corrections for the granular systems. These calculations could be performed
diagrammatically, although establishing the non-perturbative results for
both the conductivity, Eq. (\ref{b23}), and the DOS, Eq. (\ref{c8}), would
be considerably more difficult. Clearly, the action $S$ described by Eqs. (%
\ref{a9}-\ref{a11}) may not be used at zero temperature and the dissipation
resulting from this action is not a zero temperature effect.

If the tunnel conductance $g$ becomes of the order of $1$ or smaller
diagrammatic expansions are no longer helpful and calculations without the
phase functional, Eq. (\ref{a9}-\ref{a11}), are hardly possible. In this
regime a proper account of non-zero winding number $k_{{\bf i}}$ is very
important. In the next section we show how calculations can be carried out
in the limit of weak coupling constants $g\ll 1$.

\section{Weak coupling between the grains}

\subsection{Phase correlation function}

Study of the granular system described by the action $S$, Eqs. (\ref{a9}-\ref
{a7}), for an arbitrary conductance $g$ is difficult. The logarithmic
behavior, Eq. (\ref{b23}), describes the conductivity of the granular system
at sufficiently large $g\gtrsim 1$. Eqs. (\ref{c5}-\ref{c8}) are also
applicable only in this limit. Not being able to consider the model for
arbitrary $g$, we restrict ourselves with the limit of small $g\ll 1$.

We will see that the temperature dependence of both the conductivity and the
DOS becomes exponential in this limit. This means that increasing the
tunneling amplitude at a fixed $T$ we go from the almost metallic regime to
an insulating one. This can be achieved experimentally changing the coupling
between the grains while measuring at the same temperature \cite{Gerber97}.

Calculation of physical quantities at small $g\ll 1$ can be performed
expanding the functional integral in Eq. (\ref{a8}) in the tunneling part $%
S_{t}$, Eq. (\ref{a11}), of the action. When calculating conductivity the
main contribution comes again from the function $\langle X_{2}\left( \tau
\right) \rangle $ in Eqs. (\ref{a7}). As concerns the tunneling density of
states we can use as before Eq. (\ref{c3}). In the lowest order one can
completely neglect $S_{t}$ in the both formulae. Then, the calculation of
the DOS, Eq. (\ref{c3}), reduces to computation of the phase correlation
function $\Pi \left( \tau \right) $ 
\begin{equation}
\Pi \left( \tau \right) =\left\langle \exp \left( -i\left( \tilde{\phi}_{%
{\bf i}}\left( \tau \right) -\tilde{\phi}_{{\bf i}}\left( 0\right) \right)
\right) \right\rangle _{S_{c}},  \label{a16}
\end{equation}
where the phases $\tilde{\phi}_{{\bf i}}\left( \tau \right) $ are introduced
in Eq. (\ref{a5}), $S_{c}$ is given by Eq. (\ref{a10}) and the averaging
should be performed with this functional. As concerns the conductivity, we
need a slightly different correlation function $\tilde{\Pi}\left( \tau
\right) $%
\begin{equation}
\tilde{\Pi}\left( \tau \right) =\left\langle \exp \left( -i\left( \tilde{\phi%
}_{{\bf ij}}\left( \tau \right) -\tilde{\phi}_{{\bf ij}}\left( 0\right)
\right) \right) \right\rangle _{S_{c}}  \label{d1}
\end{equation}
The phase correlation function $\Pi \left( \tau \right) $ is somewhat
simpler and let us show in detail how to calculate it. A proper modification
for $\tilde{\Pi}\left( \tau \right) $ is simple.

The computation of the average in Eq. (\ref{a16}) can be performed using two
different methods. A more straightforward way of calculating is to use the
definition of $\tilde{\phi}_{i}\left( \tau \right) $, Eq. (\ref{a5}), which
allows us to represent the action $S_{c}$ as 
\begin{equation}
S_{c}=S_{c}[\phi ]+S_{c}[k],  \label{d2}
\end{equation}
\begin{equation}
S_{c}[\phi ]=\frac{T}{4}\sum_{n,{\bf i,j}}\phi _{{\bf i}n}\omega
_{n}^{2}\left( B^{-1}\right) _{{\bf ij}}\phi _{{\bf j-}n},  \label{d3}
\end{equation}
\begin{equation}
S_{c}[k]=T\pi ^{2}\sum_{{\bf ij}}k_{{\bf i}}\left( B^{-1}\right) _{{\bf ij}%
}k_{{\bf j}}  \label{d4}
\end{equation}
where 
\[
B_{{\bf ij}}=\frac{e^{2}}{2}\left( C^{-1}\right) _{{\bf ij}} 
\]
Writing Eqs. (\ref{d3}-\ref{d4}) we neglected integration over the variables 
$\rho _{{\bf i}}$ from Eq. (\ref{b7}). As we have discussed in Sec. II, in
the limit $T\gg \delta $ the main contribution comes from $\rho \sim \left(
T\delta \right) ^{1/2}$ and we can simply put in all expressions $\rho =0.$
Using Eqs. (\ref{d2}-\ref{d4}) one can carry out integration over the phase $%
\phi $ and summation over the winding numbers separately. The phase
correlation function $\Pi \left( \tau \right) $ can be written as 
\begin{equation}
\Pi \left( \tau \right) =\langle \exp (-i\left( \phi _{{\bf i}}\left( \tau
\right) -\phi _{{\bf i}}\left( 0\right) \right) \rangle _{\phi }\left\langle
\exp (-2\pi ik_{{\bf i}}\tau T)\right\rangle _{k}  \label{d5}
\end{equation}
Integrating over the phase $\phi _{{\bf i}}\left( \tau \right) $ we obtain
for $0<\tau <\beta $. 
\begin{equation}
\langle \exp (-i\left( \phi _{{\bf i}}\left( \tau \right) -\phi _{{\bf i}%
}\left( 0\right) \right) \rangle _{\phi }=\exp \left( -B_{{\bf ii}}\left(
\tau -T\tau ^{2}\right) \right) ,  \label{a19}
\end{equation}

Eq. (\ref{a19}) was used for the function $\Pi \left( \tau \right) $ in many
previous works \cite{fazio,gefen}. However, Eq. (\ref{a19}) is not the final
result because the function $\Pi \left( \tau \right) $, Eq. (\ref{d5}),
contains the second average and, in addition to integrating over $\phi _{%
{\bf i}}\left( \tau \right) ,$ one must sum over the winding numbers $k_{%
{\bf i}}.$

Calculation of the second average in Eq. (\ref{d5}) can be performed using
the Poisson formula 
\[
\sum_{k=-\infty }^{\infty }f\left( 2\pi k\right) =\frac{1}{2\pi }%
\sum_{m=-\infty }^{\infty }\int_{-\infty }^{\infty }e^{imx}f\left( x\right)
dx 
\]
for any function $f\left( x\right) $.

As a result, we rewrite the second average in Eq. (\ref{d5}) as 
\begin{equation}
\left\langle \exp (-2\pi ik_{{\bf i}}\tau T)\right\rangle
_{k}=Z_{x}^{-1}\sum_{\{m_{{\bf i}}\}}\int \exp \left( -\frac{T}{4}\sum_{{\bf %
jl}}x_{{\bf j}}\left( B^{-1}\right) _{{\bf jl}}x_{{\bf l}}+i\sum_{{\bf j}}x_{%
{\bf j}}\left( m_{{\bf j}}-\tau T\delta _{{\bf ij}}\right) \right) {\cal D}x
\label{d6}
\end{equation}
where $Z$ is the normalization factor that can be obtained putting $\tau =0$
in the integral in the second line of Eq. (\ref{d6}) and ${\cal D}x=\prod_{%
{\bf i}}dx_{{\bf i}}$. Summation over all integers $m_{{\bf i}}$ for all
grains is implied. The integration over $x_{{\bf i}}$ should be performed in
the infinite limits and the integral in Eq. (\ref{d6}) can be easily
calculated. Substituting the result of the integration into Eq. (\ref{d5})
and using Eq. (\ref{a19}) we find for the phase correlation function $\Pi
\left( \tau \right) $ 
\begin{equation}
\Pi \left( \tau \right) =Z^{-1}\exp \left( -B_{{\bf ii}}\tau \right)
\sum_{\{m_{{\bf j}}\}}\exp \left( -\sum_{{\bf k}}2\tau m_{{\bf k}}B_{{\bf ki}%
}-\beta \sum_{{\bf k,l}}B_{{\bf kl}}m_{{\bf k}}m_{{\bf l}}\right)  \label{d7}
\end{equation}
where all $m_{{\bf j}}$ are integers and $Z$ is the normalization
coefficient ($\Pi \left( 0\right) =1$). The quantity $B_{{\bf ii}}$ in Eq. (%
\ref{d7}) is the charging energy of an extra electron in a grain ${\bf i}$
in an otherwise neutral system without excitations. The necessary
periodicity in $\tau $ of the function $\Pi \left( \tau \right) $ with the
period $\beta $ is evident from Eq.(\ref{d7}). The final result for the
phase correlation function $\Pi \left( \tau \right) $, Eq. (\ref{d7}), is
essentially different from Eq. (\ref{a19}) obtained by neglecting the
contribution of non-zero winding numbers. The exponent contains only linear
in $\tau $ terms, which contrasts Eq. (\ref{a19}).

Actually, the form of Eq. (\ref{d7}) is absolutely natural and can be
obtained using the standard quantum mechanical formalism instead of
calculating the functional integrals in Eq. (\ref{a16}). Within this
formalism velocities in the action in the functional integral should be
replaced by the corresponding momentum operators in the Hamiltonian. So,
instead of having the derivatives $\partial \phi /\partial \tau $ in the
action $S_{c}$, Eq. (\ref{a10}) one would have to write the operators $%
\partial /\partial \phi $. The presence of winding numbers in the action
introduces periodicity. As a result, the corresponding angle variables $\phi 
$ in the Hamiltonian formulation should be taken in the interval $[0,2\pi ]$
and all wave functions of the Hamiltonian must be periodic in $\phi $ with
the period $2\pi $.

As the result the calculation of the phase correlation function $\Pi \left(
\tau \right) $, Eq. (\ref{a16}), reduces to calculation of quantum
mechanical averages with the effective Hamiltonian $\hat{H}_{{\rm eff}}$

\begin{equation}
\hat{H}_{{\rm eff}}=\sum_{{\bf ij}}B_{{\bf ij}}\hat{\rho}_{{\bf i}}\hat{\rho}%
_{{\bf j}}\text{, \ \ \ }\hat{\rho}_{{\bf i}}=-i\partial /\partial \phi _{%
{\bf i}},  \label{a21}
\end{equation}
where the ``angles'' $\phi _{{\bf i}}$ vary between $0$ and $2\pi $. The
correlation function $\Pi \left( \tau \right) $ can be written in the form 
\begin{equation}
\Pi \left( \tau \right) =\langle e^{-i\left( \hat{\phi}_{{\bf i}}\left( \tau
\right) -\hat{\phi}_{{\bf i}}\left( 0\right) \right) }\rangle _{\hat{H}_{%
{\rm eff}}}\text{, \ }\hat{\phi}_{{\bf i}}\left( \tau \right) =e^{\hat{H}_{%
{\rm eff}}\tau }\phi _{{\bf i}}e^{-\hat{H}_{{\rm eff}}\tau }  \label{d8}
\end{equation}

In order to calculate the average with $\hat{H}_{{\rm eff}}$ in Eq. (\ref{d8}%
) one should find eigenfunctions $\Psi _{\{n\}}$of the Hamiltonian $\hat{H}_{%
{\rm eff}}$, Eq. (\ref{a21}). These eigenfunctions have the simple form 
\begin{equation}
\Psi _{\{n\}}=\prod_{{\bf i}}\exp \left( in_{{\bf i}}\phi _{{\bf i}}\right)
\label{d9}
\end{equation}
where $n_{{\bf i}}$ are integers. Calculating the matrix elements entering
Eq. (\ref{d8}) and performing summation over all states with the weight $%
\exp \left( -E_{\{n\}}/T\right) $, where $E_{\{n\}}$ are eigenenergies of
the Hamiltonian $\hat{H}_{{\rm eff}}$, Eq. (\ref{a21}), we come to Eq. (\ref
{d7}).

The operators $\hat{\rho}_{{\bf i}}$ and $\phi _{{\bf i}}$ are conjugate to
each other with the commutator 
\[
\lbrack \hat{\rho}_{{\bf i}},\phi _{{\bf i}}]=-i 
\]
Therefore we can alternatively write the operators $\hat{\phi}_{{\bf i}}$ as 
$\hat{\phi}_{{\bf i}}=\partial /\partial \rho _{{\bf i}}$. The eigenvalues
of the operator $\rho _{{\bf i}}$ are integers. The operator $\exp \left(
\pm i\hat{\phi}\right) $ acts as follows 
\begin{equation}
\exp \left( \pm i\widehat{\phi }\right) f\left( \rho \right) =\exp \left(
\mp \partial /\partial \rho \right) f\left( \rho \right) =f\left( \rho \pm
1\right)  \label{d10}
\end{equation}
This gives another convenient way of calculation of the quantum mechanical
averages.

The use of the Hamiltonian formalism for calculation of functional integrals
over the phase $\phi $ of the superconducting order parameter has been
suggested in an earlier work on granulated superconductors \cite{efetov80}.
Within this approach the effective Hamiltonian $\hat{H}_{{\rm eff}}$, Eq. (%
\ref{a21}), was derived for the operator $\rho $ of the number of Cooper
pairs and the phase correlation function $\Pi \left( \tau \right) $, Eq. (%
\ref{d7}) has been obtained. As concerns the normal metals the correct form
of the phase correlation function $\Pi \left( \tau \right) $, Eq. (\ref{d7})
has been written for the first time in our previous paper \cite{et}.

The present consideration demonstrates explicitly that accounting for the
winding numbers leads to the charge quantization. The function $\tilde{\Pi}%
\left( \tau \right) $, Eq. (\ref{d1}) can be calculated in the same way and
can be written as 
\begin{equation}
\tilde{\Pi}\left( \tau \right) =Z^{-1}\exp \left( -B_{{\bf i}}^{{\bf a}}\tau
\right) \sum_{\{m_{{\bf k}}\}}\exp \left( -\sum_{{\bf k}}2\tau m_{{\bf k}%
}\left( B_{{\bf k,i+a}}-B_{{\bf ki}}\right) -\beta \sum_{{\bf j,l}}B_{{\bf jl%
}}m_{{\bf j}}m_{{\bf l}}\right)  \label{d11}
\end{equation}
where $B_{{\bf i}}^{{\bf a}}=B_{{\bf ii}}+B_{{\bf i+a,i+a}}-B_{{\bf i,i+a}%
}-B_{{\bf i+a,i}}$ is the energy of an electron-hole excitation between
neighboring grains, ${\bf a}={\bf j-i}$. As the array of the grains is
regular we omit the subscripts ${\bf i,j}$ when writing the correlation
functions $\Pi \left( \tau \right) $ and $\tilde{\Pi}\left( \tau \right) $.

The functions $\Pi \left( \tau \right) $ and $\tilde{\Pi}\left( \tau \right) 
$ allow us to calculate the tunneling density of states and the conductivity
using Eqs. (\ref{c3}) and (\ref{a7}).

\subsection{Tunnelling density of states at $g\ll 1$}

Substituting Eq. (\ref{d7}) into Eq. (\ref{c3}) we have to calculate the
remaining integral over $\tau $ and perform the analytical continuation $%
\varepsilon _{n}\rightarrow -i\varepsilon +\delta $. It is convenient to
shift the contour of the integration and integrate first along the line $%
\left( 0,i\infty \right) $, then along $\left( i\infty ,i\infty +\beta
\right) ,$ and finally along $\left( i\infty +\beta ,\beta \right) $. The
integral over the second segment vanishes and we reduce the function $\tilde{%
\nu}_{{\bf i}}\left( \varepsilon _{n}\right) $, Eq. (\ref{c3}), to the form 
\begin{eqnarray}
\tilde{\nu}_{{\bf i}}\left( \varepsilon \right) &=&i\nu
_{0}[1-2T\int_{0}^{\infty }dt\frac{\exp \left( -\varepsilon _{n}t\right) }{%
\sinh \left( \pi Tt\right) }\sin \left( tB_{{\bf ii}}\right)  \label{d12} \\
&&\times Z^{-1}\sum_{\left\{ m_{{\bf i}}\right\} }\exp (-\beta \sum_{{\bf kl}%
}m_{{\bf k}}B_{{\bf kl}}m_{{\bf l}}-2it\sum_{{\bf k}}m_{{\bf k}}B_{{\bf ki}%
})]  \nonumber
\end{eqnarray}
Now the analytical continuation can be done easily. Taking the imaginary
part of the function $\tilde{\nu}$ according to Eq. (\ref{c2}) and using the
integral 
\[
\int_{0}^{\infty }\frac{\sin au}{\sinh \pi u}du=\frac{1}{2}\tanh \frac{a}{2} 
\]
we obtain for the density of states $\nu \left( \varepsilon \right) $%
\begin{eqnarray}
\frac{\nu \left( \varepsilon \right) }{\nu _{0}} &=&Z^{-1}\sum_{\{m_{{\bf k}%
}\}}\exp \left( -\frac{\sum_{{\bf kl}}m_{{\bf k}}B_{{\bf kl}}m_{{\bf l}}}{T}%
\right)  \label{d13} \\
&&\times \left( n\left( \frac{\varepsilon +B_{{\bf ii}}-2\sum_{{\bf k}}m_{%
{\bf k}}B_{{\bf ki}}}{T}\right) +n\left( \frac{-\varepsilon +B_{{\bf ii}%
}+2\sum_{{\bf k}}m_{{\bf k}}B_{{\bf ki}}}{T}\right) \right)  \nonumber
\end{eqnarray}
where 
\begin{equation}
Z=\sum_{\{m_{{\bf k}}\}}\exp \left( -\frac{\sum_{{\bf kl}}m_{{\bf k}}B_{{\bf %
kl}}m_{{\bf l}}}{T}\right)  \label{d13a}
\end{equation}
and 
\[
n\left( x\right) =\frac{1}{e^{x}+1} 
\]
is the Fermi distribution function. In Eqs. (\ref{d13}, \ref{d13a})
summation over all positive and negative integer $m_{{\bf k}}$ should be
performed.

The function $\nu \left( \varepsilon \right) $ is even in the energy $%
\varepsilon $ and approaches $1$ in the limits $T\rightarrow \infty $ or $%
\left| \varepsilon \right| \rightarrow \infty $. At low temperatures $T\ll
B_{{\bf ii}},\left| \varepsilon \right| $ and $\left| \varepsilon \right|
<B_{{\bf ii}}$ the main contribution comes from the ground state
configuration when all $m_{{\bf i}}=0$. In this limit we obtain 
\begin{equation}
\frac{\nu \left( \varepsilon \right) }{\nu _{0}}=2\exp \left( -\frac{B}{T}%
\right) \cosh \frac{\varepsilon }{T}  \label{d14}
\end{equation}
where $B=B_{{\bf ii}}$. Eq. (\ref{d14}) demonstrates that there is a gap in
the density of states and this gap is equal to the single electron charging
energy of the grain. Of course, Eq. (\ref{d13}) corresponds to a fixed
chemical potential in the grain, which means that the grain is not
completely isolated and there are processes that keep the chemical potential
fixed.

One can easily generalize Eq. (\ref{d13}) to the case when an additional
voltage is present in each grain. Writing the effective chemical potential $%
\mu _{{\bf i}}$ of a grain ${\bf i}$ as 
\[
\mu _{{\bf i}}=\mu +2\sum_{{\bf j}}N_{{\bf j}}B_{{\bf ij}} 
\]
we can generalize Eq. (\ref{d13}) by replacing everywhere $m_{{\bf k}%
}\rightarrow m_{{\bf k}}+N_{{\bf k}}$. If $N_{{\bf k}}$ is randomly
distributed over the grains one should carry out an additional averaging
over this variable. This is beyond the scope of the present paper.

To the best of our knowledge, Eq. (\ref{d13}) has not been written before,
although it clearly follows from the ``orthodox theory'' of the Coulomb
blockade \cite{likh}. The latter can be seen from the fact that the one
particle part $\hat{H}_{0}$ of the Hamiltonian, Eq. (\ref{a2}) commutes with
the term $\hat{H}_{c}$, Eq. (\ref{a3}), describing the charging energy. This
means that in the absence of tunneling between the grains the charging
energy should be simply added to the Fermi energy for non-interacting
particles. Eq. (\ref{d13}) clearly corresponds to this picture. For an
arbitrary configuration $\{m_{{\bf k}}\}$ of charges on the grains the
energy of adding one electron (hole) to a grain is equal to 
\[
B\pm 2\sum_{{\bf k}}m_{{\bf k}}B_{{\bf ki}}
\]
We see that just this energy enters the Fermi distribution functions in Eq. (%
\ref{d13}) shifting the Fermi energy. The exponential in Eq. (\ref{d13}) is
the weight for the configuration of the charges in the system and two Fermi
functions describe contributions of electrons and holes. Although the direct
derivation ``in the electron language'' would be simpler, the present
calculation demonstrates explicitly how the phase functional $S_{c}$, Eq. (%
\ref{a10}), works. We emphasize that without the summation over the winding
numbers the correct result could not be (and has not been) obtained.

\subsection{Conductivity at $g\ll 1$}

Calculation for the conductivity can be performed in the same way as for the
tunnelling density of states. It is important that in the limit $\ g\ll 1$%
the main contribution comes from the term $X_{2}$ in Eq. (\ref{a7}). The
contribution coming from the term $X_{1}$ is of higher order in $g$ and we
neglect it. Then, the Fourier transformed response function $K\left( \Omega
_{n}\right) $, Eq. (\ref{a7}), takes the form 
\begin{equation}
K\left( \Omega _{n}\right) =2\pi e^{2}gT^{2}\int_{0}^{\beta }d\tau \frac{%
1-\exp \left( i\Omega _{n}\tau \right) }{\sin ^{2}\pi T\tau }\tilde{\Pi}%
\left( \tau \right)  \label{d15}
\end{equation}
where, again, the principal value of the integral should be taken.

Shifting the contour of integration over $\tau $ in the same way as it has
been done when calculating the tunnelling density of states we can perform
the analytical continuation $\Omega _{n}\rightarrow -i\omega +\delta $. In
the limit of small frequencies $\omega $ we expand the integral in $\omega .$
The first non-vanishing term is linear in $\omega $ and we reduce the
formulae for the conductivity $\sigma $ using Eqs. (\ref{a7}, \ref{a12}) to
the form 
\begin{equation}
\sigma =\sigma _{0}\left( 1-2\pi T\int_{0}^{\infty }dt\frac{Tt}{\sinh
^{2}\left( \pi Tt\right) }\sin \left( tB_{{\bf i}}^{{\bf a}}\right) {\cal C}%
_{{\bf i}}^{{\bf a}}\left( t\right) \right)  \label{d16}
\end{equation}
where 
\begin{equation}
{\cal C}_{{\bf i}}^{{\bf a}}\left( t\right) =Z^{-1}\sum_{\{m_{{\bf k}%
}\}}\exp \left( -\beta \sum_{{\bf kl}}m_{{\bf k}}B_{{\bf kl}}m_{{\bf l}%
}\right) \cos \left( 2t\sum_{{\bf k}}m_{{\bf k}}\left( B_{{\bf k,i+a}}-B_{%
{\bf ki}}\right) \right) ,  \label{d17}
\end{equation}
$Z$ is given by Eq. (\ref{d13a}) and $\sigma _{0}$ is the classical
conductivity, Eq. (\ref{a12}). Calculation of the integral over $t$ in Eq. (%
\ref{d16}) can easily be performed using the formula 
\begin{equation}
\int_{0}^{\infty }du\frac{u\sin \left( au\right) }{\sinh ^{2}\left( \pi
u\right) }=\frac{1}{2\pi }\left( 1-f\left( a\right) \right) ,  \label{d18}
\end{equation}
\[
f\left( a\right) =2\frac{\left( a-1\right) e^{-a}+e^{-2a}}{\left(
1-e^{-a}\right) ^{2}} 
\]
and we obtain finally 
\begin{equation}
\sigma =\sigma _{0}Z^{-1}\sum_{\{m_{{\bf k}}\}}\exp \left( -\beta \sum_{{\bf %
kl}}m_{{\bf k}}B_{{\bf kl}}m_{{\bf l}}\right) f\left( \beta (B_{{\bf i}}^{%
{\bf a}}+2\sum_{{\bf k}}m_{{\bf k}}\left( B_{{\bf k,i+a}}-B_{{\bf ki}%
}\right) )\right)  \label{d20}
\end{equation}

We see from Eq. (\ref{d20}) that, in order to get an explicit expression for
the conductivity, one should sum again over all charge configurations (In
Eq. (\ref{d20}) the conductivity is calculated between the grains ${\bf i}$
and ${\bf i+a}$).

The limit of high temperatures $T$ exceeding the charging energies $B$ can
be obtained using the property $f\left( 0\right) \rightarrow 1$ when $%
a\rightarrow 0$. In this limit we come to $\sigma =\sigma _{0}$, which
demonstrates that as soon as the Coulomb energy is not important the
transport is described by the Drude formula.

In the opposite limit, $T\ll B$, the main contribution in Eq. (\ref{d20})
comes from charge configurations with the lowest charging energies. For the
calculation of the conductivity we need also the asymptotics of the function 
$f\left( a\right) $ in the limit $a\rightarrow \infty $. In this limit we
may write this function as 
\begin{equation}
f\left( a\right) \simeq 2a\exp \left( -a\right)  \label{d21}
\end{equation}

As when calculating the tunnelling density of states, we consider first the
contribution of the ground state configuration with all $m_{{\bf i}}=0$.
However, in contrast to the tunnelling density of states, this configuration
gives not necessarily the main contribution because it contains a large
energy $B_{{\bf i}}^{{\bf a}}$ of a dipole consisting of an additional
electron in the grain ${\bf i}$ and a hole in the grain ${\bf i+a}$ (or vice
versa). The energy of this dipole is equal to $2B$ if we neglect
non-diagonal $B_{{\bf ij}}.$ If non-diagonal components $B_{{\bf ij}}$ are
not equal to zero the dipole energy is smaller but we assume that it is
larger than the charging energy of one electron $B$. Physically, this
contribution corresponds to transport in a completely neutral grains. In
order to contribute to the current an electron must jump from one grain to
another. However, this costs an energy which is just the dipole energy $B_{%
{\bf i}}^{{\bf a}}$. So, the contribution of the ground state configuration
can be estimated as $\exp \left( -B_{{\bf i}}^{{\bf a}}/T\right) $ and we
want to show that a larger contribution exists.

The most efficient process contributing to the current is when an additional
charge exists in a grain but all other grains are neutral. Then, jumping
from grain to grain costs no energy. As such a configuration is not the
ground state, the probability to have this state is proportional to $\exp
\left( -B/T\right) $. However, the overall contribution to the conductivity
is in this case larger because $B<B_{{\bf i}}^{{\bf a}}$.

This picture clearly follows from Eq. (\ref{d20}). Two configurations give
the main contribution to $\sigma $ in Eq. (\ref{d20}). We can put $m_{{\bf k}%
}=1$ at ${\bf k=i}$ and $m_{{\bf k}}=0$ for all ${\bf k\neq i}$ or $m_{{\bf k%
}}=-1$ for ${\bf k=i+a}$ and $m_{{\bf k}}=0$ for all ${\bf k\neq i+a}$. In
both the cases the argument of the function $f$ in Eq. (\ref{d20}) is equal
to zero and we obtain for the conductivity at low temperatures

\begin{equation}
\sigma =2\sigma _{0}\exp \left( -B/T\right) .  \label{a22}
\end{equation}

Eq. (\ref{a22}) shows that in the limit of small couplings between the
grains $g$ the macroscopic conductivity is exponentially small in
temperature. This is a typical example of an activation process.

Of course, the exponential behavior of the physical quantities, Eqs. (\ref
{d14}, \ref{a22}), was derived under the assumption that all the grains are
mesoscopically equal (they have the same size and shape but may have small
irregularities different for different grains). In real samples the shape
and the size may vary and qualitative estimates show that instead of the
activation law a dependence of the type $\exp \left( -A/\sqrt{T}\right) $
can be more proper for this case \cite{abeles}.

\section{Discussion}

We studied effects of the Coulomb interaction on the conductivity $\sigma $
and tunneling density of states $\nu \left( \varepsilon \right) $ of
granular metals. Calculations with the Hamiltonian for interacting electrons
were reduced to calculation of functional integrals with a phase action of a
form proposed by Ambegaokar, Eckern Sch\"{o}n\cite{aes}. This action has
been derived recently for the granular systems microscopically \cite{belalt}%
, which allowed to clarify conditions for its applicability. These
conditions, Eqs. (\ref{ad1}, \ref{ad2}), correspond to the limit of not very
low temperatures, such that weak localization effects are suppressed. In the
limit of large tunnelling conductances $g$ the results obtained with AES
functional correspond to Altshuler-Aronov corrections \cite{altar} and could
be calculated diagrammatically (although this way of calculations would be
more complicated). At smaller $g$ non-zero winding numbers become very
important and we developed a proper scheme of calculations.

Although the interference effects leading to localization corrections are
neglected at such temperatures, interesting effects occur. In the limit of
large $g$ a logarithmic dependence of the conductivity $\sigma $ on
temperature, Eq. (\ref{b18}), is obtained. Eqs. (\ref{b18}, \ref{b23}) are
applicable in any dimension and at any magnetic field, which distinguishes
it from the weak localization correction. The logarithm in Eqs. (\ref{b18}, 
\ref{b23}) is not just a small correction and these formulae are applicable
until the conductance becomes of order unity (It is important that the
conductivity and not the resistivity is linear in the logarithm of the
temperature).

In contrast to the conductivity $\sigma $, the tunnelling density of states $%
\nu \left( \varepsilon \right) $ is dependent on the dimensionality $d$ of
the system in a non-trivial way in the limit $g\gg 1$. In $3D$ it has a
logarithmic correction, Eq. (\ref{c5}), but is described by a more
complicated formula, Eq. (\ref{c8}), in $2D$. This formula is well known for
disordered films \cite{fin,andreev} where it was obtained within the $\sigma 
$-model approach. We see that the same formula is valid for the granular
metal at not very low temperatures and it can be obtained without $\sigma $%
-models.

In the limit of a low coupling between the grains both the tunnelling
density of states and the conductivity are exponentially small, Eqs. (\ref
{d14}, \ref{a22}). This is due to a finite charging energy arising when the
electron tunnels from one to another grain. Although the results in the
limit of vanishing $g$ are rather simple, it was important to derive them
from the AES phase functional with a proper summation over the winding
numbers. This way of calculations is very close to the one suggested
previously for granular superconductors \cite{efetov80}. In some experiments 
\cite{Gerber97} the dependence of the conductivity on temperature is
described not by the activation type formulae but by the function $\exp
\left( -a/\sqrt{T}\right) .$ This dependence may originate from fluctuations
of the charging energy of the grains \cite{abeles}.

Comparing Eqs. (\ref{b18}, \ref{c5}, \ref{c8}) obtained in the limit $g\gg 1$
with Eqs. (\ref{d14}, \ref{a22}) derived for $g\ll 1$ we conclude that there
must be a considerable change of the temperature behavior of the physical
quantities when changing the coupling between the grains. Of course, this
cannot be a sharp transition because we consider the limit of finite
temperatures and cannot extrapolate the results to $T=0$. However, tuning
the coupling experimentally at a given temperature one may see the change of
the regimes that would look like a ``metal-insulator'' transition. We
emphasize that this ``transition'' should occur in any dimension of the
array of the grains.

It is not clear from the present consideration whether there should be a
sharp transition from the metallic to the insulator state at a critical
value $g_{c}$ and this is a definitely interesting problem for a further
investigation. This problem is closely related to the question whether the
activation energy (Coulomb gap) turns to zero or has a jump at $g=g_{c}$. \ 

The model of the granular metal may be used to describe disordered electron
systems at low electron density. In such systems electrons can spend a
considerable time in traps or ``puddles'' that can be due to strong
fluctuations of a disorder potential. Considering such systems with the
model of a granular metal may be a reasonable approximation. In this case,
potential wells where electrons are trapped would correspond to the grains
in the model of the granular metal.

The model considered in the present paper and the results obtained can be
relevant to many experiments on different materials. Of course, specially
prepared granular metals like those considered in Refs. \cite
{Gerber97,abeles,simon} should be the first object of the application of the
theory developed. As the model and the results obtained are quite robust,
one can expect that the corresponding phenomena have been observed in the
granular materials.

This is really so and many experimental data can be explained in the
framework of our model. First, let us make a comparison of our theory with
experimental results of Ref. \cite{Gerber97} on films made of $Al$ grains
embedded in an amorphous $Ge$ matrix. At low temperature superconductivity
of in $Al$ grains was destroyed by a strong magnetic field. Depending on the
coupling between the grains (extracted from the conductivity at room
temperatures) the samples of the experiment \cite{Gerber97} were
macroscopically either in an insulating state with the temperature
dependence of the resistivity $R\sim \exp \left( a/T^{1/2}\right) $ or in a
``metallic'' one. However, the resistivity of the metallic state depended on
temperature and the authors suggested Eq. (\ref{a1}) to describe this
dependence. As the exponent $\alpha $ for the ``metallic'' sample was small
we may argue that Eq. (\ref{a1a}) should not be worse for fitting the
experimental data. Then, we can estimate the exponent $\alpha $ without
using fitting parameters.

The sample of the experiment \cite{Gerber97} had the room temperature
resistivity $R_{0}=7.3\times 10^{-3}\Omega $cm. The diameter of the grains
was $120\pm 20$\AA $,$ which allows, using the value $\hbar /e^{2}=4.1\times
10^{3}\Omega $, to estimate the dimensionless tunnel conductivity as $g=0.7$%
. If we put $d=2$ in Eq. (\ref{b13a}) we obtain $\alpha =0.116$, which would
perfectly agree with the experimental value $\alpha =0.117$ from Eq. (\ref
{a1}). However, everything is not so simple because the films used in Ref. 
\cite{Gerber97} were rather thick and, at first glance, one should use $d=3$%
. This would change the result by $30\%$ making the agreement less exciting.
Nevertheless, the value of $d$ in Eq. (\ref{b13a}) corresponds rather to the
half of the contacts of a single grain than to the real dimensionality.
Therefore the experimental value of $\alpha $ indicates that either the
grains are not closely packed such that the typical number of contacts per
grain is $4$ or our calculation is too rough to provide a quantitative
agreement with the experiment (the value of $\alpha $, Eq. (\ref{b13a}), is
based on the assumption $g\gg 1$ but the experimental value of $g$ is of
order $1$).

The resistivity of samples with a high room temperature resistivity (a weak
coupling between the grains) behaved as $\exp \left( a/T^{1/2}\right) $
rather than obeying the activation law, Eq. (\ref{a22}). According to Ref. 
\cite{abeles} this can be attributed to a variation of the size of the
grains or of the local potential. However, a model considered in Ref. \cite
{abeles} is rather special because the distance between the grains and the
tunnelling amplitude was assumed to be related to the charging energy in a
certain way. The law $\exp \left( a/T^{1/2}\right) $ is rather common for
granular materials with weak coupling (see also e.g. Ref. \cite{simon}) and
the reason for such an universality is still not clear. A random hopping
conduction mechanism of Ref. \cite{efros} suggested for semiconductors can
hardly be used for the granular metals.

A logarithmic dependence of the resistivity on temperature has been observed
in other granular materials. In Ref. \cite{simon} a granular cermet
consisting of $NbN$ grains in a boron nitride insulating matrix was studied.
Again, at small coupling between the grains the temperature dependence of
the resistivity $\exp \left( a/T^{1/2}\right) $ was observed in a very broad
interval of temperatures. The resistivity of samples with a strong coupling
between the grains was very well described by the law 
\begin{equation}
R=R_{0}\ln \left( T_{0}/T\right)  \label{k1}
\end{equation}
which is close to Eqs. (\ref{a1}, \ref{a1a}) if the temperature interval is
not very large such that the variation of the resistivity is small. However,
the law, Eq. (\ref{k1}), gave a good description for the temperature
dependence of the resistivity in a very broad region and the changing of the
resistivity was not small. The reason for the applicability of Eq. (\ref{k1}%
) in a so broad interval of temperatures is not clear because according to
the results of the renormalization group analysis of Sec. III not the
resistivity but the conductivity should obey Eq. (\ref{k1}). A more careful
experimental study might clarify this question. Anyway, the logarithmic
behavior of Refs. \cite{Gerber97,simon} remained unexplained at all and our
work is the first attempt to construct a theory of this effect (an
explanation in terms of weak localization corrections or the Kondo effect
can be excluded immediately because the logarithmic temperature dependence
was observed also in very strong magnetic fields and the systems were three
dimensional).

The unusual logarithmic behavior of the type, Eq. (\ref{k1}), has been
observed not only in ``standard'' granular systems but also in high-$T_{c}$
cuprates at very strong magnetic fields. The first observation of this
dependence was done on underdoped $La_{2-x}Sr_{x}CuO_{4}$ crystals\cite{boeb}%
. The superconductivity in this experiment was suppressed with pulsed
magnetic fields of $61T$. It was found that both the in-plane resistivity $%
\rho _{ab}$ and out-of-plane resistivity $\rho _{c}$ diverged
logarithmically with decreasing the temperature. This means again that a $3D$
effect was observed in a very strong magnetic field and traditional
explanations like localization or Kondo effect could not clarify the
situation.

In a subsequent publication \cite{boeb1} a metal-insulator crossover was
observed in the same material at a $Sr$ concentration near optimum doping ($%
x\simeq 0.16$). In underdoped samples both $\rho _{ab}$ and $\rho _{c}$
showed no evidence of saturation at low temperatures and diverged as the
logarithm of the temperature. The authors called this state ``insulator'' in
contrast to the state at high doping where the resistivity did not have a
pronounced dependence on the temperature. It was conjectured in Ref. \cite
{boeb1} that the logarithmic behavior they observed might be related to the
one seen in the experiment \cite{simon} on granular $NbN$. This would demand
a phase segregation throughout the underdoped regime of $LSCO$. However, as
no explanation had been given for the logarithmic behavior in Ref. \cite
{simon}, no explanation has been given to the experiments \cite{boeb,boeb1}
either. We hope that our results for the model of the granular materials may
be applicable to the experiments on the $La_{2-x}Sr_{x}CuO_{4}$ crystals 
\cite{boeb,boeb1}, which would mean that the underdoped crystals have a
granular structure and the logarithmic behavior is due to the Coulomb
interaction. The transition to the metallic state of Refs. \cite{boeb,boeb1}
would mean that at higher doping the granularity disappears.

The logarithmic dependence of the resistivity on temeperature has also been
observed in many other experiments. For example, in Ref. \cite{gerber90}
this dependence was observed in granular $Pb$ films. It was also observed in
phase compounds of $Nd_{2-x}Ce_{x}CuO_{4-y}$, Ref.\cite{radha}. In each case
the reason for such a behavior was not clear.

An interesting conclusion has been made recently about the structure of a $%
2D $ gas in $GaAs/AlGaAs$ heterostructures where a metal-insulator
transition was observed \cite{yacoby}. Measuring the local electronic
compressibility the authors found that the metallic phase was homogeneous in
space, which is natural. In contrast, the system becomes spatially
inhomogeneous as it crosses into the insulating phase. The structure seen in
the insulating state indicates that the electrons were located in
``puddles''. So, modelling the system in terms of a granular metal might be
a reasonable approximation also in this case.

We are grateful to A. Altland and B.L. Altshuler, A.V. Andreev, L.I.
Glazman, A. Gerber, A.I. Larkin, and Yu.V. Nazarov for discussions. A
support of the SFB/Transregio 6029 and German-Israeli Foundations (GIF and
DIP) is greatly appreciated.


\end{document}